\documentclass[a4paper,11pt]{article}

\usepackage{geometry}
\usepackage{graphicx}
\usepackage{float}
\usepackage{subcaption}
\usepackage{multirow}
\usepackage{ragged2e}
\usepackage{blindtext}

\usepackage{amsmath}
\usepackage{amssymb}
\usepackage{slashed}
\usepackage{tikz-feynman}
\tikzfeynmanset{compat=1.1.0} 

\usepackage{authblk}  
\usepackage{ulem}     

\usepackage[style=numeric-comp, sorting=none, backend=biber]{biblatex}
\usepackage{hyperref}

\title{\textbf{Disappearing Track Signals from a Light Charged Higgs in the Alternative Left-Right Model}}

\author[1]{\textbf{Hrishikesh Deka}\thanks{\texttt{hrishikesh.deka@iitg.ac.in}}}
\author[2]{\textbf{Avnish}\thanks{\texttt{avnishy@uohyd.ac.in}}}
\author[3]{\textbf{Poulose Poulose}\thanks{\texttt{poulose@sju.edu.in}}}

\affil[1]{Department of Physics, Indian Institute of Technology Guwahati, Assam - 781039, India.}
\affil[2]{School of Physics, University of Hyderabad, Hyderabad - 500046, India.}
\affil[3]{Department of Physics, St. Joseph's University, Bengaluru - 560 027, India.}

\date{} 

\begin{document}

\maketitle


\begin{abstract}
We study the phenomenology of a light charged Higgs boson in the framework of the Alternative Left--Right Symmetric Model (ALRM). In this model, stringent flavor constraints are evaded due to a non-conventional fermion spectrum in which right-handed up-type quarks are paired with exotic down-type quarks rather than the Standard Model down-type quarks, leading to the absence of tree-level flavor-changing neutral currents. Furthermore, a specific assignment of the global $U(1)_S$ symmetry and the resulting emergent $R$-parity prevent mixing between the right- and left-handed charged gauge bosons, $W_R$ and $W_L$, providing additional suppression of flavor-violating effects. The ALRM accommodates potentially viable dark matter candidates, both fermionic and scalar ones. In this context, an associated charged Higgs state, $H_2^\pm$, belonging to the dark sector can naturally acquire a sub-TeV to TeV-scale mass without conflicting with any experimental constraints. We focus on scenarios in which $H_2^\pm$ behaves as a long-lived particle due to a sub-GeV mass splitting with the dark matter candidate. We identify regions of parameter space consistent with the observed dark matter relic density and other experimental constraints. A detailed analysis of disappearing track signatures is performed, including realistic tracklet reconstruction efficiencies, and the existing ATLAS search are recast to assess the current limits and future sensitivities. We find that the HL-LHC has limited sensitivity to TeV-scale charged Higgs bosons in this scenario, while the 27~TeV HE-LHC can effectively probe the relevant parameter space, with a 100~TeV collider offering substantially enhanced discovery potential.
\end{abstract}

\vspace{0.5cm} 

\section{Introduction}

Despite its envious success, the Standard Model (SM) of particle physics leaves several fundamental questions unanswered. Among these, the nature of dark matter, the origin of neutrino masses, and the observed baryon asymmetry of the universe are the most prominent ones that point towards the incompleteness of the SM framework. In addition to these phenomenological shortcomings, the SM also suffers from internal theoretical limitations that motivate physics beyond it. The foremost among these is the issue of vacuum stability: for the measured values of the Higgs and top-quark masses, the SM Higgs potential develops an instability at high scales, indicating that the electroweak vacuum is only metastable. Among the best approaches to addressing this instability is the introduction of a richer Higgs sector, which can modify the running of scalar couplings, stabilize the vacuum, and simultaneously accommodate some of the other shortcomings of the SM. Interestingly, such extensions are often also motivated by deeper structural issues of the Standard Model, one of which is the essentially ad hoc implementation of maximal parity violation in the weak interactions.
Although this feature is phenomenologically required, it is highly desirable to understand it as an outcome of a more fundamental symmetry. The left–right symmetric framework was introduced precisely for this purpose, restoring parity at high energies by enlarging the gauge symmetry to include a right-handed analogue of the electroweak interaction. In this setup, the right-handed quarks form doublets under an additional $SU(2)_R$ symmetry, and an analogous structure is introduced in the leptonic sector by pairing each right-handed charged lepton with a right-handed neutrino. This constitutes the paradigm of the standard Left–Right Symmetric Model (LRSM) \cite{PhysRevD.11.2558, PhysRevD.12.1502}. In the LRSM, parity violation emerges naturally through the spontaneous breaking of the right-handed gauge symmetry at a scale above the electroweak scale. The Higgs sector is required to be much larger than the single scalar field present in the SM. An added bonus of this framework is the automatic incorporation of the seesaw mechanism, which accounts for the observed tiny  neutrino masses \cite{PhysRevD.11.2558, PhysRevD.17.2462}. However, this comes at a price: the minimal LRSM contains tree-level flavour-changing neutral currents (FCNCs) \cite{Maiezza:2010ic, ECKER1983365}, which in turn require the $SU(2)_R$  breaking scale to be pushed to very high values to satisfy experimental limits. As a result, many of the new particles predicted by the model become too heavy to be accessible at current collider energies. Moreover, the conventional LRSM lacked a dark matter candidate, although it could address the matter-antimatter asymmetry problem to some extent. 

An alternative approach to the left–right symmetric model (ALRM) was introduced to address this difficulty in Ref. \cite{Ma:1986we, Babu:1987kp}. Part of the motivation arises from the observation that the gauge sector of the LRSM can be embedded in grand unified scenarios such as the exceptional group \(E_6\), having
\(\,SU(3)\times SU(3)\times SU(3)\,\) as one of its maximal subgroups. This can break further to \(\,SU(3)\times \, SU(2)\times SU(2)\times U(1)\,\), yielding the right gauge group required for left-right symmetric frameworks. When the particle content is considered in this context, it is natural to take the fermions to belong to the 27-dimensional fundamental representation of $E_6$, which accommodates all Standard Model fermions along with additional exotic states.
In the conventional LRSM, the breaking pattern is chosen such that the right-handed up-type and down-type quarks form an $SU(2)$ doublet. Similarly, the right-handed electron is paired with the right-handed neutrino to make an $SU(2)$ doublet. However, this is not the only possible assignment. The ALRM exploits an alternative pairing: the right-handed up-type quark is paired with an exotic right-handed down-type quark, forming an $SU(2)_R$ doublet, while the right-handed charged lepton is paired with an exotic right-handed neutral fermion, not the usual seesaw right-handed neutrino, to form the corresponding leptonic doublet. In this scheme, parity symmetry is not introduced as a fundamental assumption of the low-energy theory but is instead subtly encoded in the underlying $SU(3)\times SU(3)\times SU(3)$ structure, from which the left–right gauge symmetry emerges after symmetry breaking.
A global $U(1)_S$ symmetry is introduced in addition to the gauge symmetry described above. As we shall elaborate in the next section, assigning appropriate $U(1)_S$ charges to the various fields 
forbids FCNC interactions at tree-level.
Furthermore, while \( U(1)_S \) is broken simultaneously with \( SU(2)_R \), a residual \( Z_2 \) symmetry remains, enabling the assignment of \( R \)-parity to all particles. With an appropriate choice of \( U(1)_S \) charges, the neutral scalar fields (\( H_1^0 \) and \( A_1^0 \)), exotic quark fields (\( d' \)), and the neutral exotic fermion fields (\( n \)) become \( R \)-parity odd, whereas all Standard Model particles are \( R \)-parity even. This structure ensures the stability of potential dark matter candidates.
The global \(U(1)_S\) also forbids the mixing between \(W_L\) and \(W_R\). Combined with the altered fermion content of the model, this renders the standard LHC search limits on \(W_R\) inapplicable. In contrast, the neutral gauge boson \(Z_R\) mixes with the electroweak gauge boson \(Z_L\), so the heavy neutral states are directly constrained by LHC searches. Finally, through the mass relation between \(W_R\) and \(Z_R\), these bounds translate into an indirect lower limit on the mass of \(W_R\)\cite{Frank:2019nid}.
 
A similar situation arises for one of the charged Higgs bosons, denoted as \(H_2^\pm\). Being \(R\)-parity odd, it decays via non-standard fermionic channels, making conventional collider search constraints irrelevant for this state. These features are exploited in the phenomenological analysis of \(W_R\) and \(H_2^\pm\) production at the LHC and its high-luminosity upgrade, the HL-LHC, as well as the proposed high energy version, HE-LHC \cite{Frank:2024imi}.  
Belonging to the dark sector, together with potential dark-matter candidates,  \(H_1^0/A_1^0\) and  \(n\), the charged scalar \(H_2^\pm\) can give rise to exotic collider signatures. In particular, in regions of parameter space where the mass splitting between \(H_2^\pm\) and the dark matter candidate is small, its decay width is suppressed, and \(H_2^\pm\) can behave as a long-lived particle (LLP).  For sufficiently small mass splitting, $H_2^\pm$ might decay into dark matter and very soft pions, resulting in disappearing track signatures at the collider experiments. In the recent past,  disappearing track signatures had been studied in the context of the super-symmetry (SUSY) \cite{Ostdiek:2015aga, Low:2014cba, Mahbubani:2017gjh, Fukuda:2017jmk, Bharucha:2018pfu, Filimonova:2018qdc, Saito:2019rtg, Fukuda:2019kbp, Han:2018wus, }, type-III seesaw \cite{Jana:2020qzn}, and other scenarios \cite{Belyaev:2020wok,Chiang:2020rcv, Belanger:2022gqc,Bandyopadhyay:2024plc}. These disappearing track signatures have been recently explored by the ATLAS \cite{ATLAS:2017oal, ATLAS:2022rme}, and the CMS \cite{CMS:2018rea, CMS:2020atg} collaborations. In addition, studies of disappearing track signals at the lepton colliders are presented in the literature \cite{Kadota:2018lrt, Capdevilla:2021fmj}.

In this work, we explore the aforementioned possibilities in detail, identifying regions of parameter space consistent with current constraints, including those from the dark-matter sector and LHC searches in the context of the ALRM framework. We also provide projections for how future experiments could further probe and constrain this parameter space. In Section~\ref{model_framework}, we provide some details of the model framework. In the following sections, we focus on the phenomenology of \(H_2^\pm\): in Sec.~\ref{DM}, we summarize the properties of the dark-sector particles arising from the \(R\)-parity odd sector, and in Sec.~\ref{H+inLHC}, we analyze the production and decay profiles of \(H_2^\pm\), considering two scenarios in which it decays either to a scalar or a fermionic dark-matter candidate. Finally, our results are summarized and conclusions presented in Sec.~\ref{conclusion}.

\section{The Model Framework}\label{model_framework}

The ALRM is another variant of the left-right symmetric model  based on the symmetry group
\begin{center}
${\rm SU}(3)_{\rm C} \otimes {\rm SU}(2)_{\rm L} \otimes {\rm SU}(2)_{{\rm R^{'}}} \otimes {\rm U}(1)_{{\rm B-L}} \otimes { U(1)_S}$,
\end{center}
where $U(1)_S$ denotes a global symmetry. The particle content of the ALRM with their respective quantum numbers is listed in Table \ref{fields}.
As in the SM, all left-handed fermions transform as doublets under $\rm SU(2)_L$. In contrast, the right-handed quarks $u_R$ and the right-handed charged leptons $e_R$ transform as doublets under  $\rm{SU(2)}_{R^\prime}$, pairing with the exotic right-handed quarks $d^\prime_{R}$ and the right-handed neutral leptons $n_R$, respectively. On the other hand, the left-handed exotic quarks ($d^\prime_{L}$), the right-handed down-type quarks ($d_{R}$), and the two newly introduced neutral fermions, $n_L$ and $\nu_R$, are all singlets under both $\rm{SU(2)}_L$, and $\rm{SU(2)}_{R^\prime}$ \cite{Ashry:2013loa, Frank:2019nid}. Masses of the up-type quarks and the charged leptons require the presence of a scalar bi-doublet field, $\Phi$, which transforms as $(2,2^*)$ under $\rm SU(2)_L\times SU(2)_{R'}$. On the other hand, mass terms of the down-type quarks do not arise from the bi-doublet as in the conventional LRSM. Instead, the presence of $\chi_L$, which transforms as a doublet under $\rm SU(2)_L$ and as a singlet under $\rm SU(2)_{R'}$, and its Yukawa interaction with the left- and the right-handed down-type quarks leads to their masses. The extended gauge symmetry is broken to the Standard Model gauge group through the non-zero vacuum expectation value (vev) of $\chi_R$, which is charged under $\rm SU(2)_{R'}$ but singlet under $\rm SU(2)_L$.

\begin{table}[]
 \centering
 \begin{tabular}{ c | c | c | c | c | c||c |c} 
		\hline \hline
		Particles  &${\rm SU}(3)_{\rm C}$  &${\rm SU}(2)_{\rm L}$  &${\rm SU}(2)_{{\rm R'}}$  &${\rm U}(1)_{{\rm B-L}}$   &$U(1)_{\rm S}$ 
        &$L$&$R$\\
       \hline
		
       $Q_L$ = $\begin{pmatrix}
		u_L \\ 
		d_L 
		\end{pmatrix}$ &3 &2 &1 &$\frac{1}{6}$ &0&0&+ \\ [5mm]
        
       $Q_R$=$\begin{pmatrix}
		u_R \\ 
		d^{\prime}_{R}
		\end{pmatrix}$ &3 &1 &2 &$\frac{1}{6}$ & - $\frac{1}{2}$&$\begin{pmatrix}
		0 \\ 
		-1
		\end{pmatrix}$&$\begin{pmatrix}
		+ \\ 
		-
		\end{pmatrix}$\\[5mm] 
       
       $d^{\prime}_{L}$    &3 &1 &1 & - $\frac{1}{3}$ &$-1$&$-1$&$-$  \\ 
       
       $d_{R}$        &3 &1 &1 &- $\frac{1}{3}$ &0&0&+ \\  
       \hline
        
       $L_L$ = $\begin{pmatrix}
		\nu_L \\ 
		e_L 
		\end{pmatrix}$  &1 &2 &1 &-$\frac{1}{2}$ & 1&1&+\\  [5mm]  
       
       $L_R$ = $\begin{pmatrix}   
		n_R \\ 
		e_R 
		\end{pmatrix}$  &1 &1 &2 &-$\frac{1}{2}$ &  $\frac{3}{2}$&$\begin{pmatrix}   
		2 \\ 
		1 
		\end{pmatrix}$ &$\begin{pmatrix}   
		- \\ 
		+ 
		\end{pmatrix}$ \\
        
       $n_L$           &1 &1 &1 &0 &2&2&$-$\\
        
       $\nu_R$         &1 &1 &1 &0 &1&1&+\\
		\hline  
       
       $\Phi$ = $\begin{pmatrix}
	    \phi_1^{0}  & \phi_1^{+} \\ 
	    \phi_2^{-}  & \phi_2^{0}
	    \end{pmatrix}$ &1 &2 &$2^*$ &0 & -$\frac{1}{2}$
        &$\begin{pmatrix}
	    -1  & 0 \\ 
	    -1 & 0
	    \end{pmatrix}$&	$\begin{pmatrix}
	    -  & + \\ 
	    -  & +
	    \end{pmatrix}$ \\[5mm]
		
       $\chi_L$=$\begin{pmatrix}
	    \chi_L^{+} \\ 
	      \chi_L^{0}
	    \end{pmatrix}$  &1 &2 &1 &$\frac{1}{2}$ &0&0&$\begin{pmatrix}
	    + \\ 
	     +
	    \end{pmatrix}$  \\[5mm]
	    
       $\chi_R$=$\begin{pmatrix}
	    \chi_R^{+} \\ 
	    \chi_R^{0}
	    \end{pmatrix}$  &1 &1 &2 &$\frac{1}{2}$ &$\frac{1}{2}$&$\begin{pmatrix}
	    1 \\ 
	    0
	    \end{pmatrix}$ &$\begin{pmatrix}
	    - \\ 
	     +
	    \end{pmatrix}$  \\
    \hline\hline
   \end{tabular}
   \caption{The particle content of ALRM with their respective gauge structure. The generalised lepton number is defined as $L=S+T_{R'3}$, and the $R$-parity as $R=(-1)^{3B+L+2s}$, where $B$ is the baryon number and $s$ the spin.}
   \label{fields}
\end{table}

The most general Yukawa interaction part of the Lagrangian can be written as  

\begin{equation}\label{Yukawa}
\begin{split}
      	- \mathcal{L}_Y =  Y^{q}_{ij} ~\overline{Q}_{L_i}  \tilde{\Phi} Q_{R_j} + Y^{q}_{L_{ij}} \overline{Q}_{L_i} \chi_L d_{R_j}  + Y^q_{R_{ij}} \overline{Q}_{R_i} \chi_R d^{'}_{L_j} + Y^{\ell}_{ij} ~\overline{L}_{L_i} \Phi L_{R_j} \\
       + Y^\ell_{L_{ij}} \overline{L}_{Li} \tilde{\chi}_L\nu_{R_j}  + Y^\ell_{R_{ij}} \overline{L}_{R_i} \tilde{\chi}_R n_{L_j} + M_{N_{ij}} \overline{\nu}^c_{R_i} \nu_{R_j} + h.c. ,~~~~~
\end{split}        
\end{equation}
where $\tilde \Phi$=$\sigma_2 \Phi^* \sigma_2$ and $\tilde \chi_{L,R}$=$i\sigma_2 \chi_{L,R}$ are the duals of respective fields with $\sigma_2$ being the second type Pauli matrix. The indices $i,j=1,2,3$ are the family indices.
 In an advantage over conventional LRSM, at the tree-level, we are allowed to introduce a $U(1)_S$ soft-breaking Majorana mass term $M_{N_{ij}} \overline{\nu}^c_{R_i} \nu_{R_j}$ to the right-handed neutrinos for being gauge singlets. Turning to the scalar sector, the most general scalar potential $V_{\Phi\chi}$ respecting the symmetry structure of the ALRM is given by

\begin{equation}\label{alrpot}
\begin{split}
 V_{\Phi \chi} & = - \mu_1^2 {~\rm Tr} \left[\Phi^{\dagger}\Phi \right] - \mu_2^2 \left(\chi_L^{\dagger}\chi_L +\chi_R^{\dagger}\chi_R \right) +   \lambda_1 \left( {\rm Tr} \left[\Phi^{\dagger}\Phi \right] \right)^2 +             \lambda_2 ~{\rm Tr} \left[\Phi^{\dagger}\tilde{\Phi} \right]           {\rm Tr} \left[\tilde{\Phi^{\dagger}}
     \Phi \right] \\
     & +  \lambda_3 \left[ \left(\chi_L^{\dagger}\chi_L \right)^2 + \left(\chi_R^{\dagger}\chi_R \right)^2 \right] 
      +2\lambda_4 \left(\chi_L^{\dagger}\chi_L \right) \left(\chi_R^{\dagger}\chi_R \right) +2\alpha_1 {\rm Tr} \left[\Phi^{\dagger}\Phi \right] \left(\chi_L^{\dagger}\chi_L+\chi_R^{\dagger}\chi_R \right)\\
      &+ 2\alpha_2 \left[\chi_L^{\dagger}\Phi\Phi^{\dagger}\chi_L+\chi_R^{\dagger}\Phi^{\dagger}\Phi\chi_R \right] +2\alpha_3 \left[\chi_L^{\dagger}\tilde{\Phi} {\tilde{\Phi}}^{\dagger} \chi_L +\chi_R^{\dagger} {\tilde{\Phi}}^{\dagger} \tilde{\Phi}\chi_R \right] \\
     &  +\mu_3 \left[\chi_L^{\dagger}\Phi\chi_R+\chi_R^{\dagger} \Phi^{\dagger} \chi_L \right],
 \end{split}
 \end{equation}
where we treat all couplings to be real for simplicity. The detailed properties of this potential and the implications for the vacuum stability of the model have been addressed in Ref. \cite{Frank:2021ekj}. 

In the broken symmetry phase we consider the following assignment of vacuum expectation values (vev) for the respective scalar fields:
\begin{equation}    
\langle\Phi\rangle=\frac{1}{\sqrt{2}}
\begin{pmatrix}
    0 & 0 \\
    0 & v_u
\end{pmatrix},~~~~~~
\langle \chi_{L}\rangle=\frac{1}{\sqrt{2}}
\begin{pmatrix}
    0  \\
     v_{L}
\end{pmatrix},~~~~~~
{\rm and}~~ \langle \chi_{R}\rangle=\frac{1}{\sqrt{2}}
\begin{pmatrix}
    0  \\
     v_{R}
\end{pmatrix}.
\end{equation}
The vev of the right-handed doublet $v_R$ breaks the gauge group $SU(2)_{R'}\times U(1)_{B-L}$  to the standard hypercharge symmetry group $U(1)_Y$. 
At the same time, the global $U(1)_S$ is spontaneously broken to a discrete subgroup $Z_2$ , which allows the assignment of an $R$-parity to all particles. The $R$-parity is defined as $R = (-1)^{3B+L+2s}$, where $B$ is the baryon number and $s$ is the spin of the particle considered, while $L$ is a generalized lepton number defined as $L=S+T_{R'3}$, with S as the $U(1)_S$ charge and $T_{R'3}$ as  the third component of the $\rm SU(2)_{R'}$ isospin. The resulting 
$R$-parity is not imposed ad hoc, but arises as the residual $Z_2$ symmetry left unbroken (but for the soft-breaking Majorana mass term) after the spontaneous breaking of the global $U(1)_S$ symmetry by the right-handed scalar vev. The vev of $\Phi$ ($v_u$) and the $\chi_L$ ($v_L$) break the standard electroweak symmetry group down to $U(1)_{em}$ as required. The direction of the vev of $\Phi$ is chosen so that one of the neutral component of $\Phi$, the $\phi_1^0$ does not acquire vev, while the other one, $\phi_2^0$ does.  Notably, $\phi_1^0$ is odd under the above $R$-parity, whereas $\phi_2^0$ and $\chi_L^0$ are even. Thus, this choice of vev direction preserves the residual $Z_2$ symmetry, ensuring the stability of the $R$-odd particles. 

It is worth noting that $W_R$ is odd under the $R$-parity, disallowing it to mix with $W_L$. 
The masses of the charged gauge bosons are given by
\begin{equation}
    M_{W_L} = \frac{1}{2}g_L \sqrt{v_u^2+v_L^2} \equiv \frac{1}{2}g_Lv ~~~~ {\rm and} ~~~~ M_{W_R} = \frac{1}{2}g_R \sqrt{v_u^2+v_R^2} \equiv \frac{1}{2}g_Rv',
\end{equation}
where we have the standard vev, $v=\sqrt{v_u^2+v_L^2}$  and $v'=\sqrt{v_u^2+v_R^2}$. Here, $g_L$ and $g_R$ refer to the coupling constants of $\text{SU(2)}_{\text{L}}$ and $\text{SU(2)}_{\text{R'}}$ interactions, respectively. Since $W_R$ is odd under $R$-parity, it does not contribute to the low-energy observables, in particular  flavor-violating processes such as $K^0-\overline{K}^0$ and $B^0-\overline{B}^0$ mixing. This effectively removes all flavor sector constrains on its mass. Moreover, $W_R$ cannot decay into the Standard Model final states, rendering the usual LHC direct-search constraints inapplicable. The neutral gauge bosons, on the other hand can mix to give the mass eigenstates, the usual $Z$ boson and the heavy $Z'$ with masses given by \cite{Frank:2024bss}
\begin{align}
    M_Z^2 =  \frac{v^2}{4} \left(g_L^2+g_Y^2\right), ~~~~ {\rm and} ~~~~
    M_{Z'}^2 = \frac{g^4_{BL} v^2+g^4_R v_u^2+v_R^2(g_{BL}^2+g_R^2)^2}{4\left(g_{BL}^2+g_R^2\right)}.
\end{align}
where $g_{BL}$ and $g_Y$ are the gauge coupling constants of $\text{U(1)}_{\text{B-L}}$ and $\text{U(1)}_{\text{Y}}$ gauge groups, respectively.

The masses of fermions are generated as usual through Yukawa interactions. For the Standard Model fermions, the masses are given by
\begin{equation}    
    m_u= \frac{1}{\sqrt{2}}Y^q v~{\text{sin}}~\beta,~~m_d= \frac{1}{\sqrt{2}}Y^q_{L} v~{\text{cos}}~\beta,~~{\rm and}~~ 
   m_\ell= \frac{1}{\sqrt{2}}Y^\ell v~ {\text{sin}}~\beta,  
\end{equation}
where tan$\beta=\frac{v_u}{v_L}$. The exotic fermions, on the other hand, receive their masses entirely from $v_R$:
\begin{equation}    
m_{d^{'}}= \frac{1}{\sqrt{2}}Y^q_{R} v_R,~~{\rm and}~~m_n= \frac{1}{\sqrt{2}}Y^\ell_{R}v_R.
\end{equation}
Similar to $W_R$, the down-type exotic quarks $d'$, and the exotic neutral fermion $n$ are both odd under the $R$-parity, ensuring that these particles do not mix with the corresponding Standard Model particles. Coming to the neutrino mass, as mentioned before, the soft breaking of $U(1)_S$ allows us to write down a Majorana mass term for $\nu_R$, as listed in Eq.~\ref{Yukawa} along with the Yukawa interaction terms.\footnote{ This could potentially generate matter-antimatter asymmetry through leptogenesis \cite{Frank:2020odd}.} Through the standard type-I seesaw mechanism \cite{Mohapatra:1979ia,Schechter:1980gr}, this generates the small neutrino masses 
\begin{equation}
m_\nu = M_\nu M^{-1}_N M^T_\nu
\end{equation}
for the light neutrinos, where \( M_\nu = \frac{1}{\sqrt{2}}Y^l_L v_L\), while the heavy neutrino masses are given by $M_N$.

Let us turn to the scalar sector. Before symmetry breaking, the scalar sector contains a total of sixteen real degrees of freedom: eight from the bi-doublet $\Phi$, four from the left-handed doublet $\chi_L$, and four from the right-handed doublet $\chi_R$. After the electroweak and the right-handed symmetry breaking, two neutral degrees of freedom become part of the neutral gauge bosons $Z$ and $Z'$, and four charged degrees of freedom are absorbed to give masses to the charged gauge bosons $W_L^\pm$ and $W_R^\pm$. Therefore, the remaining ten real degrees of freedom arrange themselves as: four charged scalars (two $H^+,~ H^-$ pairs), four CP-even neutral scalars, and two CP-odd neutral scalars.

Because $R$-parity is conserved, mixing between $R$-parity--even and $R$-parity--odd scalar states is forbidden. As a result, the charged scalar squared--mass matrix is block diagonal after the electroweak symmetry breaking, and decomposes into two independent $2\times2$ blocks corresponding to the $R$-parity--even basis $(\phi_1^+,\chi_L^+)$ and the $R$-parity--odd basis $(\phi_2^+,\chi_R^+)$. The gauge eigenstates are related to the physical charged Higgs and Goldstone states by the orthogonal rotations
\begin{equation}
\label{cosxi}
\begin{pmatrix}
\phi_1^+ \\
\chi_L^+
\end{pmatrix}
=
\begin{pmatrix}
\cos\beta & \sin\beta \\
-\sin\beta & \cos\beta
\end{pmatrix}
\begin{pmatrix}
H_1^+ \\
G_1^+
\end{pmatrix},
\qquad
\begin{pmatrix}
\phi_2^+ \\
\chi_R^+
\end{pmatrix}
=
\begin{pmatrix}
\cos\xi & \sin\xi \\
-\sin\xi & \cos\xi
\end{pmatrix}
\begin{pmatrix}
H_2^+ \\
G_2^+
\end{pmatrix}.
\end{equation}
with
\[
\tan\beta=\frac{v_u}{v_L},\qquad {\rm and}\qquad \tan\xi=\frac{v_u}{v_R}.\]
The states $G_1^\pm$ and $G_2^\pm$ are the charged Goldstone bosons which get absorbed by $W_L^\pm$ and $W_R^\pm$, respectively, while $H_1^\pm$ ($R$--parity even) and $H_2^\pm$ ($R$--parity odd) are the physical charged Higgs states. Their squared masses are \cite{Frank:2020odd, Frank:2022tbm}
\begin{align}
\label{eq:mH1}
m^2_{H_1^\pm} &= -\Big[v_u v_L(\alpha_2-\alpha_3)+ \tfrac{\mu_3 v_R}{\sqrt{2}}\Big]\frac{v^2}{v_u v_L},\\[6pt]
\label{eq:mH2}
m^2_{H_2^\pm} &= -\Big[v_u v_R(\alpha_2-\alpha_3)+ \tfrac{\mu_3 v_L}{\sqrt{2}}\Big]\frac{v'^2}{v_u v_R}, 
\end{align}
respectively, where $v^2\equiv v_u^2+v_L^2$ and $v'^2\equiv v_u^2+v_R^2$. In the special limit $\alpha_2=\alpha_3$ \footnote{As discussed in \cite{Frank:2021ekj}, $\mu_3<0$ for the pseudo-sclar $A_2$ mass to be real, and further $\alpha_2-\alpha_3$ is required to be positive for a stable vacuum. This demands $\alpha_2-\alpha_3 \ll 1$  for $m_{H_2^\pm}$ to be real. }, one obtains the simple relation
\begin{equation}
m_{H_1^\pm} \;=\; \frac{v\,v_R}{v_L\,v'}\, m_{H_2^\pm},
\end{equation}
and the parameter $\mu_3$ may be expressed in terms of $m_{H_2^\pm}$ as
\begin{equation}
\label{eq:mu3_formula}
\mu_3 \;=\; -\frac{\sqrt{2}\,v_u v_R}{v_L v'^2}\, m_{H_2^\pm}^2 .
\end{equation}

In the neutral scalar sector, $\phi_1^0$ is $R$-parity odd and therefore does not mix with the $R$-parity--even neutral states. As a consequence, its real and imaginary components remain unmixed and form two exactly mass-degenerate physical eigenstates, denoted $H_1^0$ (CP-even) and $A_1$ (CP-odd), with masses
\begin{equation}
\label{eq:mH1A1}
m^2_{H_1^0}=m^2_{A_1}= 2\lambda_2 v_u^2 -(\alpha_2-\alpha_3)(v_L^2+v_R^2) - \frac{\mu_3 v_L v_R}{\sqrt{2}v_u}.
\end{equation}
The mass degeneracy between $H_1^0$ and $A_1$ is protected by symmetry. Since $\phi_1^0$ carries odd $R$-parity and a nonzero generalized lepton number, the scalar potential depends only on $|\phi_1^0|^2$. Operators of the form $(\phi_1^0)^2+\text{h.c.}$, which would split the CP-even and CP-odd components, violate the conserved quantum numbers and are therefore forbidden. As a result, the degeneracy remains exact and is not lifted by perturbative quantum corrections. Although lepton number is violated by Majorana mass terms in the fermion sector, the scalar potential involving $\phi_1$ possesses an accidental global symmetry consistent with exact $R$-parity. As a result, operators that would split the CP-even and CP-odd components of $\phi_1^0$ are forbidden, and the degeneracy remains exact to all orders in perturbation theory. If $H_1^0$ and $A_1$ are the lightest $R$-parity odd states, they jointly serve as dark-matter components.

The real components of the remaining neutral scalars, $\phi_2^0,~\chi_L^0$ and $\chi_R^0$ mix together to have three $CP$-even scalar particles, $h,~H_2^0,~H_3^0$. We consider one of them, $h$, as the Standard Model like Higgs boson. The masses of these states are discussed in detail in Ref. \cite{Frank:2021ekj}. Whereas, the $CP$-odd states mix to give two Goldstone bosons, later forming part of $Z$ and $Z'$, and one pseudo-scalar particle, denoted as $A_2$. The mass of $A_2$ is given by \cite{Frank:2021ekj}
\begin{equation}
\label{eq:mA2}
m^2_{A_2}= -\frac{\mu_3 v_L v_R}{\sqrt{2} v_u}\left[ 1+ v_u^2\Big(\frac{1}{v_L^2}+\frac{1}{v_R^2}\Big)\right].
\end{equation}
It is the $R$-parity odd charged Higgs boson, $H_2^\pm$ that is of our interest in this study. It is an admixture of the gauge eignestates $\chi_R^\pm$ and $\phi_2^\pm$. However, for the $v_R$  considered to be around 10 TeV or more, the $\chi_R^\pm$ component is practically absent. $H_2^\pm$ has direct interaction  with the down-type exotic quark, $d'$ and the corresponding up-type quark. This Yukawa interaction is controlled by the coupling, $Y^q$. The other Yukawa interaction of $H_2^\pm$ is with the charged lepton and the exotic fermion $n$ through the coupling, $Y^\ell$. The mass relations we discussed above clearly indicates that $H_2^\pm$ is much lighter than $H_1^\pm$, the other charged Higgs in the model, and therefore, its decay involving $H_1^\pm$ is highly suppressed. The other possibility is $H_2^\pm$ decays into a scalar and the gauge bosons. Here, $H_2^\pm$ directly couples to the $R$-parity odd neutral scalars, $H_1^0/A_1$ along with $W_L$. This decay is controlled entirely by the mass of the neutral scalar. On the other hand, $H_2^\pm$ decays involving $R$-parity even neutral Higgs bosons, including the Standard Model like Higgs, is highly suppressed as it is associated with $W_R$. 

The influence of $H_2^\pm$ can affect the dark sector dynamics significantly. We shall first examine this, and find the parameter space regions connected to $H_2^\pm$ that is compatible with the dark-matter observations.

\section{Dark Matter}\label{DM}
Earlier studies investigating dark matter dynamics within the ALRM framework include Ref. \cite{ Frank:2019nid, Frank:2022tbm, Khalil:2009nb}, which explore both scalar and scotino dark matter candidates. In those works, the mass hierarchy in the dark sector was chosen such that the charged Higgs boson $H_2^\pm$ decays promptly. Consequently, the collider signatures were characterized by the production of dark-sector states followed by prompt decays, with missing transverse energy and momentum playing a central role. 
In contrast, in the present work we explore the complementary region of parameter space where $H_2^\pm$ becomes a long-lived particle (LLP). This leads to distinctive disappearing-track signatures at colliders such as the LHC, dramatically altering the search strategies and phenomenology.
For any viable dark sector, consistency with astrophysical and laboratory constraints on dark matter must be ensured. The Planck measurement of the temperature anisotropies in the cosmic microwave background (CMB) determines the total dark matter relic abundance to be $\Omega h^2 = 0.12 \pm 0.001$ \cite{Planck:2018vyg}, where $h$ is the reduced Hubble parameter. 

Experimental searches for dark matter can broadly be grouped into three categories: (a) direct detection, (b) indirect detection, and (c) collider searches.  
In the direct detection of dark matter, experiments such as XENON \cite{XENON:2018voc, XENON:2023cxc}, LUX-ZEPLIN (LZ) \cite{LZ:2024zvo}, PandaX \cite{PandaX-II:2017hlx}, and PICO \cite{PICO:2019vsc} have placed stringent limits on dark matter–nucleon scattering cross sections from the non-observation of signal events. 
In the indirect detection of dark matter, searches by Fermi-LAT \cite{Karwin:2016tsw} and H.E.S.S. \cite{Hess:2021cdp, Kerszberg:2023cup} probe dark matter annihilation or decay products in astrophysical environments. Collider experiments such as the Large Hadron Collider (LHC) provide a powerful laboratory to probe dark matter and its associated dynamics beyond the Standard Model. At hadron colliders, dark matter particles can be produced either in pairs or in association with the visible Standard Model (SM) states. Since the dark sector particles escape the detector without interactions, their production is typically inferred from a recoiling SM object, leading to the well-known mono-$X$ signatures characterized by a single jet, photon, or electroweak boson accompanied with a large missing transverse energy. Beyond these canonical missing-energy searches, a growing emphasis has been placed on more unconventional collider signatures that arise naturally in extended frameworks such as the ALRM. In particular, the presence of feebly interacting or kinematically suppressed states can give rise to long-lived particles, whose macroscopic decay lengths lead to striking experimental signatures such as displaced vertices, disappearing tracks, or metastable charged particles. These signatures are actively pursued not only at the LHC but also at the intensity-frontier experiments such as the Belle II \cite{Belle-II:2025bhd}, and provide a complementary and often more sensitive probe of the dark sector. In this context, the ALRM predicts the possibility of long-lived charged Higgs bosons, whose collider phenomenology we now proceed to discuss in detail.


For our numerical analysis, we implemented the ALRM in $\texttt{SARAH}$\cite{staub2012sarah, Staub:2015kfa} to generate the corresponding \texttt{UFO}\cite{Darme:2023jdn} model files, as well as the source code for \texttt{SPheno-4.0.5} \cite{Porod:2011nf} and \texttt{CalcHEP-3.9.2} \cite{Pukhov:2004ca}. These outputs were subsequently interfaced with \texttt{micrOmegas-6.2.3}\cite{Alguero:2023zol} to compute the dark matter relic density and to evaluate the relevant cross sections for both direct and indirect dark matter detection. For collider phenomenology, event generation was performed using \texttt{CalcHEP-3.9.2} along with \texttt{Pythia8}\cite{Bierlich:2022pfr} for the parton showering and hadronization, followed by a detector-level simulation with \texttt{Delphes-3.5.0}\cite{deFavereau:2013fsa}, adopting a LHC-like detector configuration appropriate for the Large Hadron Collider.

Rather than working directly with the Lagrangian-level parameters, it is more convenient to trade them for physical observables such as particle masses. The explicit relations defining this parameter mapping are provided in Appendix~\ref{tadpole}. Accordingly, we choose the following set of independent input parameters:
\begin{equation}
\lambda_2,~ \lambda_3,~ \alpha_{1,2,3},~ \tan\beta,~ v_R,~ v,~ m_h,~ m_{H_2}^\pm,~ m_{n_i},~ m_{d^\prime_i},
\label{inputparams}
\end{equation}
where $m_{n_i}$ denote the masses of the scotinos (neutral fermionic dark matter candidates), and $m_{d^\prime_i}$ correspond to the masses of the exotic quarks. These parameters are specified together with the right-handed gauge coupling $g_R$, which is allowed to differ from the left-handed coupling $g_L$ maintaining the generality of the asymmetric left–right framework.

We consider the scalar and the scotino dark matter scenarios separately. The corresponding mass hierarchies are chosen as follows:
\begin{enumerate}
    \item {\bf Scotino Dark Matter Case}: $m_{d^\prime} >  m_{H^+_2} >m_{A_{1}} >m_{n_e}$,~ {\rm and }   
    \item {\bf Scalar Dark Matter Case}: $m_{n_e}, m_{d^\prime} > m_{H^+_2} > m_{A_{1}}$.
\end{enumerate}
In both scenarios, the exotic quark is heavier than $H_2^\pm$, which kinematically forbids the decay of $H_2^\pm$  into the exotic quark. This mass ordering arises naturally from our focus on the long-lived particle (LLP) signature of the charged Higgs boson $H_2^\pm$, as the suppression of its dominant decay modes leads to an extended lifetime. In the following sections, we analyze these two scenarios separately.

\subsection{Scotino Dark Matter Case: $m_{d^\prime} >  m_{H^+_2} >m_{H_1^0/A_{1}} >m_{n_e}$ }

In this scenario, the lightest scotino, \( n_e \), serves as the dark matter candidate, while the remaining scotinos, \( n_\mu \) and \( n_\tau \), are taken to be heavier than \( n_e \) by a mass splitting
\begin{equation}
\Delta M_{e,(\mu/\tau)} = (0\text{--}50)\,\text{GeV}.
\end{equation}
Similarly, the \(R\)-parity--odd exotic quarks \( d^\prime \) are assumed to be much heavier than the dark matter, with their masses fixed at \(\sim 5~\text{TeV}\) in our analysis. This choice ensures that the \( d^\prime \) states effectively decouple from the dark matter dynamics.

With this prescribed mass hierarchy, the \(R\)-parity--odd charged scalar \( H_2^\pm \) predominantly decays via two-body channels into the dark matter particle and a charged lepton,
\( H_2^+ \to n_e\, e^+ \).
The masses of the neutral \(R\)-parity--odd scalars, \( H_1^0 \) and \( A_1 \), are determined by the input parameters defined in Eq.~\ref{inputparams}. Throughout our numerical study, we ensure the hierarchy
\begin{equation}
m_{H_2^\pm} > m_{H_1^0/A_1} > m_{n_e},
\end{equation}
so that \( n_e \) is indeed the lightest \(R\)-parity--odd state and hence the dark matter candidate, and furthermore that \( H_2^\pm \) decays dominantly into \( n_e \).

Neglecting the electron mass, the corresponding partial decay width is given by
\begin{equation}
\label{eq:H2decay}
\Gamma_{H_2^+ \rightarrow n_e\, e^+}
=
\frac{1}{16\pi}\,
m_{H_2^+}
\left(1-\frac{m_{n_e}^2}{m_{H_2^+}^2}\right)^2
\left[
|Y^\ell|_{11}^2 \cos^2\xi
+
|Y_R^\ell|_{11}^2 \sin^2\xi
\right].
\end{equation}

Considering the thermal dark matter scenario, where \( n_e \) remains in thermal equilibrium with the Standard Model particles during the very early Universe, its freeze-out and present-day relic density are governed by the efficiency of annihilation processes. The relevant annihilation and co-annihilation channels in our model are summarized in Table~\ref{AnniCoanni_scotino_DM}, with the dark matter mass restricted to \( m_{n_e} < 2000~\mathrm{GeV} \). Given that \( m_{H_2^\pm} > m_{n_e} \), for higher dark matter masses, the production cross section of \( p~p \rightarrow H_2^+~H_2^- \) at the LHC becomes negligibly small. Since our main focus in this study is the LLP collider signature of \( H_2^\pm \), we confine our analysis to this intermediate dark matter mass range, so that sufficiently large production of $H_2^\pm$ is possible at the LHC.

\begin{table}[H]
    \centering
    \begin{tabular}{c}
    \hline\hline
    Annihilation channels \\ 
    \hline
    $n_e \bar{n}_e\rightarrow \ell \bar{\ell}, ~ q\bar{q},~hZ,~ WW,~ hh,~ \nu \bar{\nu},~ \gamma\gamma,~ZZ$ \\ \hline \hline
    Co-annihilation channels\\ \hline 
    $n_e~H_2^\pm \rightarrow h\ell,~ Z\ell$ \\
    $n_e ~n_{\mu/\tau}\rightarrow e ~ \mu/\tau$ \\
    $H_1^0H_2^\pm\rightarrow ZW,~W\gamma,~ hW$ \\
    $H_2^+H_2^-\rightarrow WW, ~ZZ,~ \gamma\gamma,~ Z\gamma , hh,~hW,~ZW$\\
    \hline\hline 
\end{tabular}
    \caption{Annihilation and co-annihilation channels relevant for the freeze-out mechanism of $n_e$ as dark matter, with $m_{n_e}<2$ TeV and under the mass hierarchies, $m_{n_{\mu,\tau}} > m_{n_e}$ and $m_{d'}>m_{H_2^\pm}>m_{n_e}$. }
    \label{AnniCoanni_scotino_DM}
\end{table}

Using \texttt{micrOmegas}, we scan the parameter ranges given in Table~\ref{tab:NeDMscanrange} and select parameter points consistent with the observed relic density and existing dark matter direct and indirect detection bounds. 
\begin{table}[h!]
\centering
\begin{tabular}{c|c|c|c|c|c|c|c}
\hline\hline
$-\lambda_2$ & $\lambda_3$ & $\alpha_{1,2,3}$ & $\tan\beta$ & $v_R$ & $m_{H_2^\pm}$ & $m_{H_2^\pm} - m_{ne}$ & $m_{n_{\mu,\tau}} - m_{ne}$ \\
\hline
$[10^{-5},0.1]$ &
$[0.01,~0.1]$ &
$0.1$ &
$[2,~100]$ &
$[8,30]~\text{TeV}$ &
$[500~,~2500~$] &
$0-50~$ &
$0-50~$ \\
\hline\hline
\end{tabular}
\caption{Range of parameters considered for the scan using parameter scan using {\texttt{micrOmegas}}. All masses are in GeV. In addition, we fix the exotic quark masses as: $m_{d'}=3$ TeV, $m_{s'}=m_{b'}=5.5$ TeV. }
\label{tab:NeDMscanrange}
\end{table}
Remember that our focus in this study is the analysis of  LLP signature of $H_{2}^\pm$. As clear from the decay width expression in Eq.~\ref{eq:H2decay}  LLP requirement makes $m_{n_e}$ to be close to $m_{H_2^\pm}$. The mass splitting required for LHC may be sub-GeV. However, for the scan, we consider a larger range of 50 GeV. Because the parameters $\alpha_{1,2,3}$ play no role in the dark-matter dynamics, we simply set their values to 0.1 throughout our analysis.  Further, as discussed in Sec.~\ref{model_framework}, we assume $\alpha_2=\alpha_3$ all through our analysis. The range of $\tan\beta=\tfrac{v_u}{v_L}$ is decided by the smallest and largest allowed $v_L$ \big(or, equivalently by the limiting values of $v_u$ as the standard model vev is given by the combination, $v=\sqrt{v_u^2+v_L^2}$~\big). As discussed in Sec.~\ref{model_framework}, the down type quark mass is given by $v_L$ through
\(
m_{b} = \tfrac{1}{\sqrt{2}}Y_L^q v_L.
\)  
Setting the perturbative limit of $Y_L^q=\sqrt{4\pi}$, and $m_b=4.2$ GeV, we have
\[
v_L > 1.7~~{\rm GeV}.
\]
Taking a slightly more relaxed value of $v_L\approx 2.5,$ we set the upper limit of $\tan\beta = 100$. The lower limit on $\tan\beta$ should respect the up-type quark masses. Considering top quark mass, $m_t = \frac{1}{\sqrt{2}}Y^q v_u=172.5$ GeV, again setting the perturbative limit of $Y^q=\sqrt{4\pi}$, we have 
\[
v_u > 69~~{\rm GeV}. 
\]
Relaxing this to close to 220 GeV (corresponding to a $Y^q\approx 1$, so as to avoid any complications with RG runnings of the model parameters), we set the lower limit of $\tan\beta=2.$  Similarly, the lower limit of $v_R$ is set keeping in mind the direct search limits on $Z'$ and consequently $W_R$ masses, whereas the upper limit on $v_R$ beyond 20-30 TeV has no relevance on the dark matter sector. The mass of the charged Higgs $H_2^\pm$ is chosen respecting both the direct and indirect constraints from LHC searches, while ensuring that the corresponding production cross section remains sufficiently large for  collider studies. The choice of $\lambda_2$ sets the mass splitting between $M_{H_2^\pm}$ and the neutral scalar, $M_{H_1^0/A_1}$ as per Eq.~\ref{eq:mH2} and Eq.~\ref{eq:mH1A1}. For the choice of $\alpha_2=\alpha_3$ that we are considering, and with $v'\approx v_R$, 
\begin{equation}
    m^2_{H_2^\pm}-m^2_{H_1^0/A_1}= -2\lambda_2 v_u^2
\end{equation}
where $\lambda_2<0$ so that $H_{1^0/A_1} < m_{H_2^\pm}$. For close to degenerate case, $m_{H_2^\pm}\approx m_{H_1^0/A_1}$, this gives
\begin{equation}
    \Delta M(H_2^\pm,H_1^0) \approx -\frac{\lambda_2~v_u^2}{m_{H_2^\pm}}
    \label{eq:DeltaMH2MD}
\end{equation}
Keeping the mass splitting $\Delta M(H_2^\pm,H_1^0)\lesssim 10$ GeV, we set our scan range of $ (-\lambda_2) < 0.1$ in our study.

\begin{figure}[H]
\centering
\includegraphics[width=.45\textwidth]{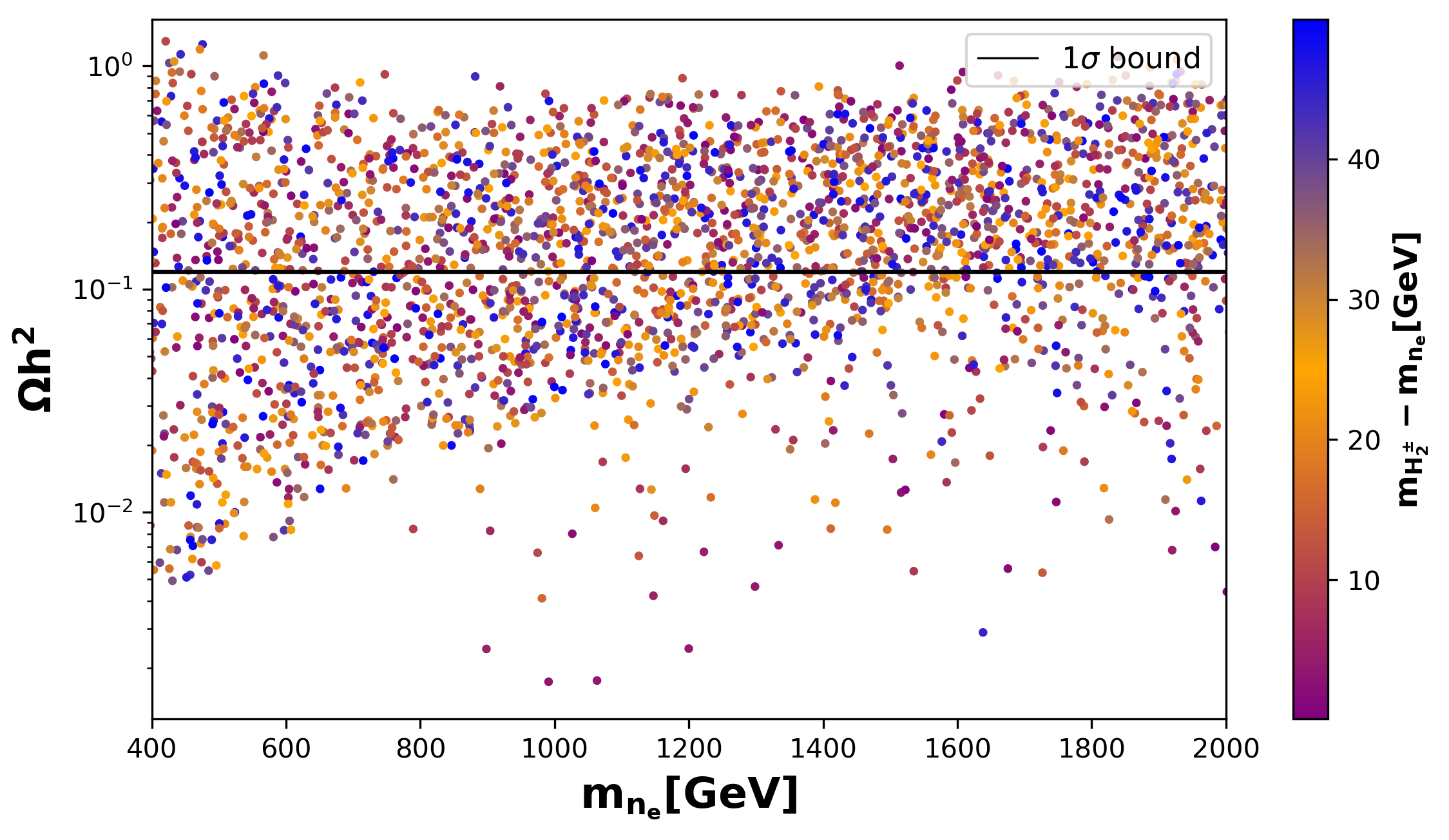}
\includegraphics[width=.45\textwidth]{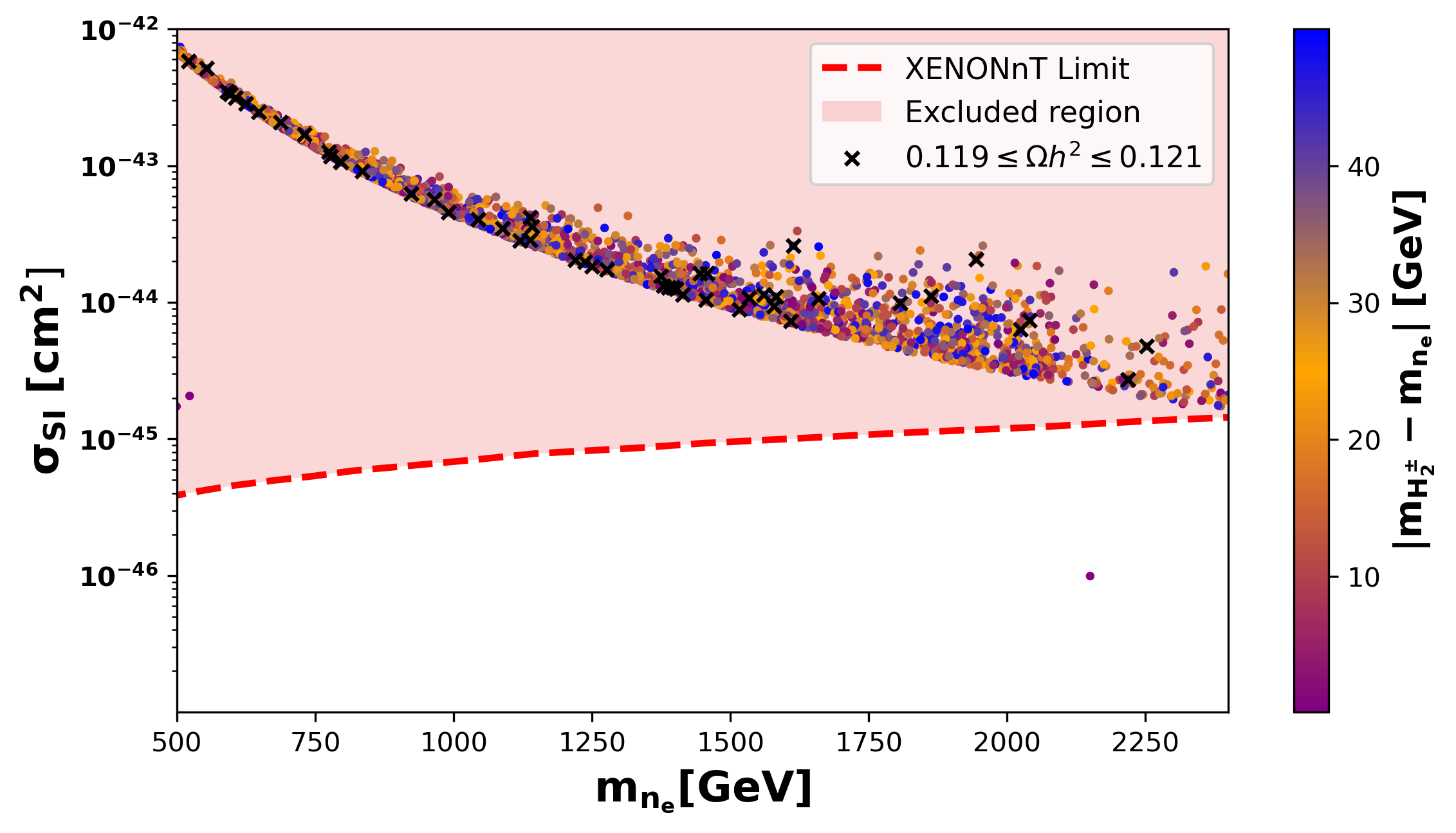}
\caption{\label{Scot_relic_mn} \textit{Left:} Variation of dark matter relic density with dark matter mass $m_{n_{e}}$. The dark black band shows the values within $1\sigma$ range of the observed relic density. \textit{Right:} Variation of direct detection cross section $\sigma_{SI}~(cm^2)$ with DM mass $m_{n_e}$. The dashed red line represents the XENONnT limit on the spin-independent direct detection cross-section of dark matter which excludes the whole region above it.  The gradient color indicates mass splitting between the dark matter and $H_2^\pm$.}
\end{figure}

We present the results of our parameter scan in Fig.~\ref{Scot_relic_mn}. The left panel illustrates the dependence of the dark matter relic density on the dark matter mass, where the gradient color scale indicates the mass splitting between the charged scalar $H_2^\pm$ and the dark matter $n_e$. The solid black band shows the \(1\sigma\) range of the observed relic density. The right panel displays the spin-independent dark matter-nucleus scattering cross-section as a function of DM mass. The dashed red line corresponds to the XENONnT \cite{XENON:2023cxc} upper limit on the spin-independent DM-nucleus cross section, which provides a strong constraint, excluding all parameter space for dark matter masses below 2.3~TeV. It remains to be seen whether scalar dark matter is a viable dark matter candidate. We shall examine this in detail in the following section.
          
\subsection{Scalar Dark Matter Scenario: $m_{n_e}, m_{d^\prime} > m_{H^+_2} > m_{A_{1}}= m_{H^0_{1}}$}
 
In this scenario, we have considered exotic quarks $d^\prime,~ s^\prime, ~b^\prime$ to be significantly massive so that it decouples from the dark matter dynamics. The relevant DM annihilation and co-annihilation channels for this scenario are presented in Table \ref{HAscalar_DM}. 

\begin{table}[H]
    \centering
    \begin{tabular}{c}
    \hline\hline
    Annihilation channels \\ 
    \hline
    $H^0_1 ~H^0_1/~ A_1~ A_1\rightarrow W~W,~ Z~Z,,~h~h,~t~\bar{t},~b~\bar{b}$ \\ \hline \hline
    Co-annihilation channels\\ \hline 
    $H^0_1 ~H_2^\pm\rightarrow Z~W,~W~\gamma$ \\
    $A_1 ~H_2^\pm\rightarrow Z~W,~W~\gamma$ \\
    $H_2^+ ~H_2^-\rightarrow W~W,~Z~Z, ~\gamma~\gamma,~Z~\gamma , ~h~h$\\
    \hline\hline
\end{tabular}
    \caption {Relevant annihilation and co-annihilation channels for $H^0_1$ and $A_{1}$ as dark matter with mass $m_{DM}<2$ TeV. Both $n_i$ and $d'_i$ are considered sufficiently heavier than $H_2^\pm$ to avoid significant co-annihilation cross sections.}
    \label{HAscalar_DM}
\end{table}
Once again, we have performed a random scan of the parameter space to identify regions consistent with current experimental constraints on the dark matter relic density and the direct detection cross-sections. The range of parameters considered for this scan is summarized in Table \ref{HAparam_mA}. 
\begin{table}[H]
\centering
\begin{tabular}{c|c}
\hline\hline
{ Parameter} & { Scan Range} \\ \hline 
 $\lambda_2$ &[$-0.1, -10^{-5} $]\\ \hline
  $\lambda_3$& [$0.01, 0.1$] \\ \hline
  $\alpha_{1,2,3}$ & 0.1 \\ \hline
  tan $\beta$&[2, 100] \\ \hline
  $v_R$& [8 - 30] TeV \\ \hline
    $m_{H_2^\pm}$ & [500 - 2000] GeV \\ \hline
    $m_{n_e}-m_{H_2^\pm}$ & [0 - 50] GeV \\ \hline
    $m_{n_\mu,n_\tau}-m_{n_e}$ & [0 - 50] GeV \\ \hline
    ($m_{d^\prime},~ m_{s^\prime}, ~m_{b^\prime}$) &($3~{\rm TeV},~ 4~{\rm TeV },~ 5~{\rm TeV}$) \\ \hline\hline
\end{tabular}
\caption{\label{HAparam_mA}The details of relevant parameter ranges considered for the scan. The dark matter mass, $M_{H_1^0/A_1}$ is very close to, but always smaller than, $M_{H_2^\pm}$ for this choice of parameter values.}
\end{table}
 
\begin{figure}[H]
\centering
\includegraphics[width=.45\textwidth]{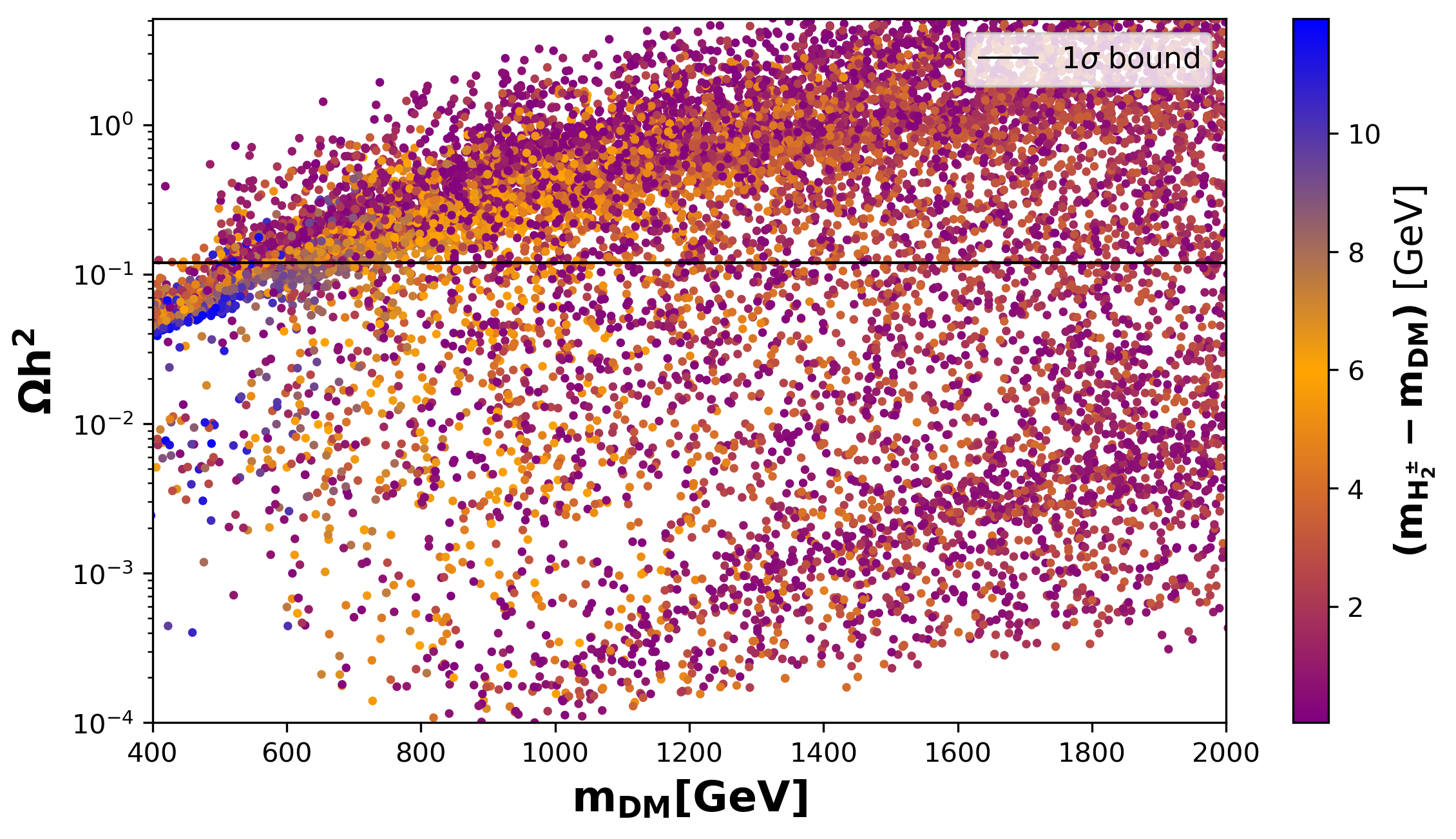}
\includegraphics[width=.45\textwidth]{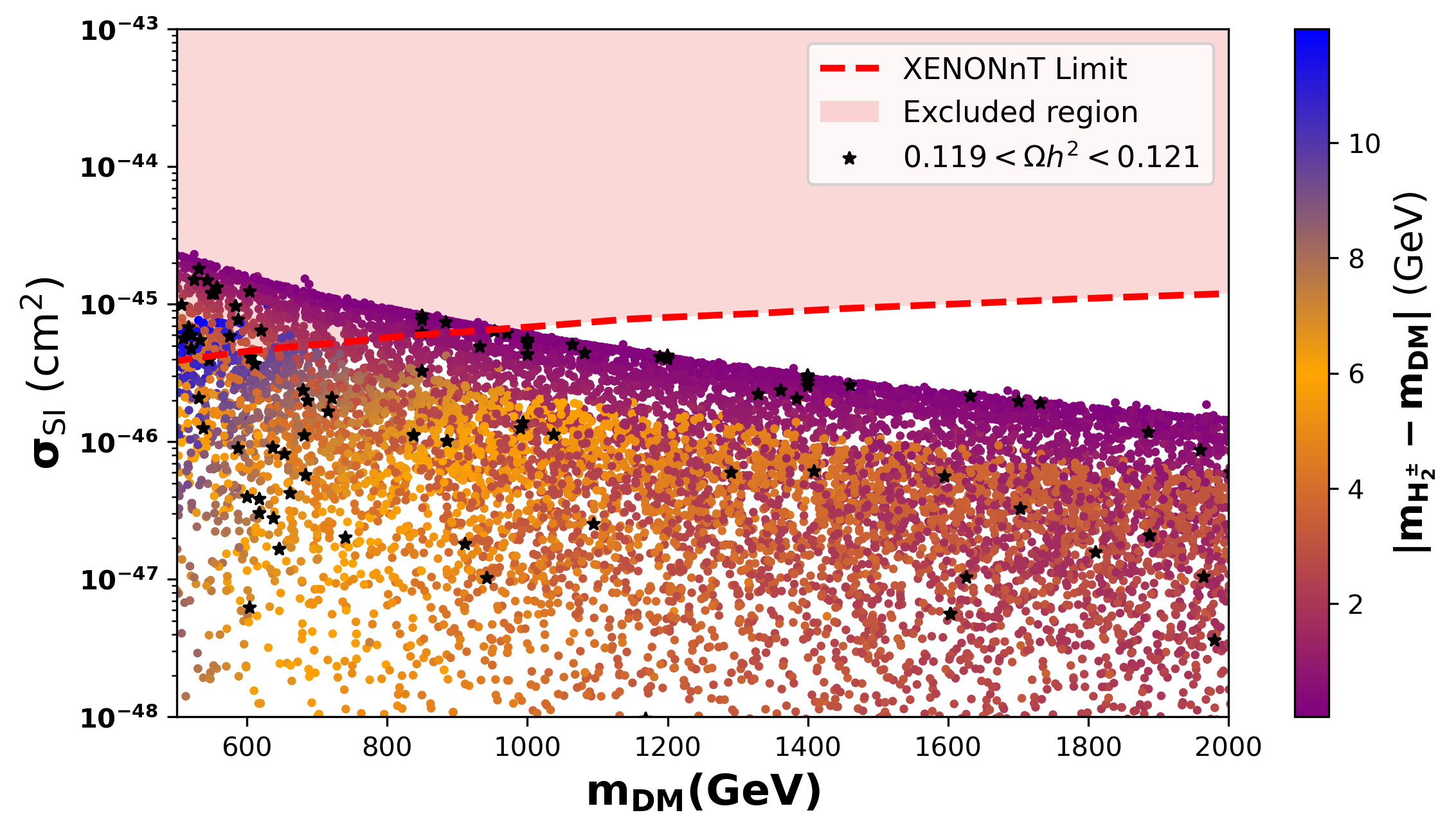}
\caption{\label{HArelic_mA_new} 
{\it Left}: The variation of darkmatter relic density with darkmatter mass $M_{H_1^0/A_1}$. The dark solid lines represent the $1\sigma$ range of the observed relic density. \\
{\it Right}: { Spin Independent cross section as a function of dark matter mass with XENONnT \cite{XENON:2023cxc} exclusion as the shaded region. The black points respect the observed relic density within $1\sigma$ limit.}\\
The color gradient denotes the mass splitting between the charged scalar $H^\pm_2$ and dark matter.
}
\label{HArelic_mA_new}
\end{figure}
The results of this scan are presented in Fig. \ref{HArelic_mA_new}. 
In the left panel of Fig. \ref{HArelic_mA_new}, we demonstrate the dark matter relic density as a function of dark matter mass and the mass splitting between the charged scalar $H^\pm_2$ and DM. Here, the dark solid lines represent the 1$\sigma$ limit of the observed relic density with the region above it being excluded. Unlike in the case of scotino dark matter, here $m_{DM} = m_{H_1^0/A_1}$ is not the free parameter. Instead we considered $\lambda_2$ as the free parameter, with $m_{DM}$ depending on $\lambda_2$ and $m_{H_2^\pm}$. For the range of $\lambda_2$ considered, $\Delta M=m_{H_2^\pm}-m_{DM}$ depends on the value of $m_{H_2^\pm}$ as per Eq.~\ref{eq:DeltaMH2MD}. For $m_{DM}=400$ GeV this gives $(\Delta M)_{max}\sim 15$ GeV, which is reduced to $(\Delta M)_{max}\sim 10$ GeV for $m_{DM}=600$ GeV, to $(\Delta M)_{max}\sim 6$ GeV for $m_{DM}=1000$ GeV and further to $(\Delta M)_{max}\sim 3$ GeV for $m_{DM}=2000$ GeV (see also Fig.~\ref{fig:DeltaM}). Thus, for example, the blue colored points are available only between 400-600 GeV of $M_{DM}$. Notice that  sub-TeV dark matter  requires large mass splitting with the charged scalar $H^\pm_2$ so that the $H_2^+H_2^-$ annihilation is relevant for dark matter relic density.
 In the right panel of Fig. \ref{HArelic_mA_new}, the spin-independent direct detection cross-section is plotted against the dark matter  mass, where the dashed  red line denotes the XENONnT limit \cite{XENON:2023cxc} on the spin-independent direct detection cross-section of dark matter with the region above being excluded. Mass above a TeV is consistent with this direct detection constraint, while for lighter scalar dark matter, larger mass splitting is required.

In summary, the observed dark matter relic density and experimental bound on $\sigma_{\rm SI}$ can be accommodated for a large region of parameter space of ALRM, featuring a sub-TeV to TeV scale dark matter and a closely mass-degenerate $H_2^\pm$. At the collider experiments, dark matter particles escape the detector and contribute to large missing energy and missing transverse momentum. In contrast, the charged scalar $H^\pm_2$  may provide detectable signatures. In particular, we shall consider the disappearing track signature of a long-lived $H_2^\pm$.

\section{Collider Studies: The charged scalar $H^\pm_2$ at the LHC}\label{H+inLHC}

\begin{figure}[H]
 \centering 
\includegraphics[width=.48\textwidth]{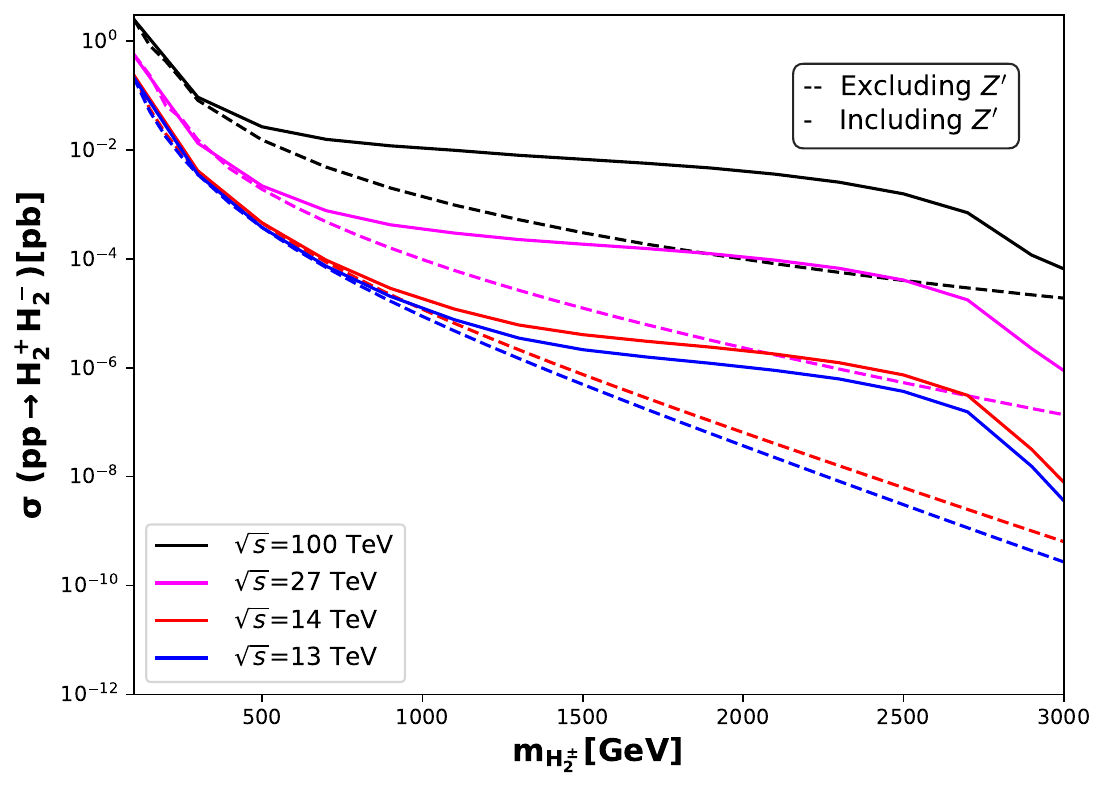}
\includegraphics[width=.48\textwidth]{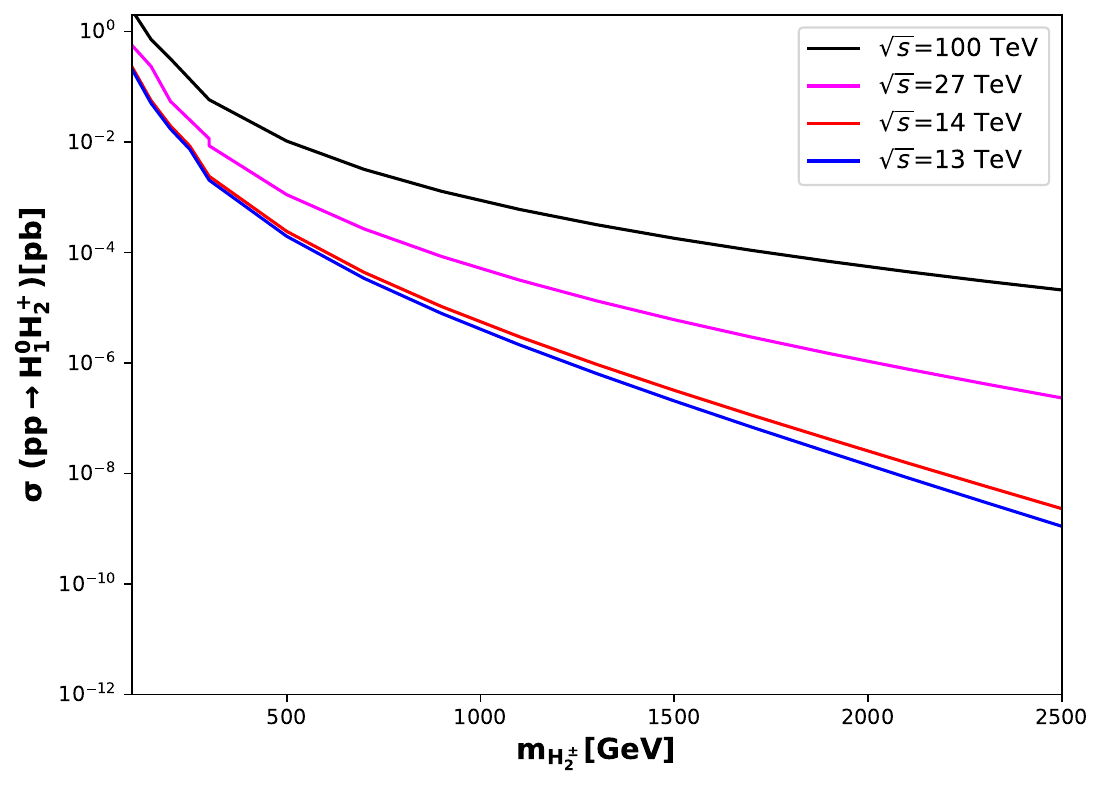}
 \caption{\label{HAProduction} \textit{Left:} Variation of the pair production cross-section of $H_2^\pm$ with its mass at the LHC.  \textit{Right}: Variation of the associated production cross section of $H_2^\pm$ along with $A_1/H^0_1$ with its mass $m_{H^\pm_2}$. To generate these plots, we have considered $v_R$ = 15 TeV, tan$~\beta$=10, $\lambda_2=-0.01$, $\lambda_3=0.01$, ($m_{d'},m_{s'},m_{b'}$)=(3 TeV, 4 TeV, 5 TeV) and $m_{Z^\prime}=5.87$ TeV.}
\end{figure}

The charged scalar $H^\pm_2$ 
can be produced in pairs and in association with the pseudo-scalar $H_1^0/A_1$ or $W^\mp_R$ at the LHC. Though the associated production with the pseudo-scalar is comparable with the pair production, the associated production with $W_R$ is very much suppressed in cross-section, as $W_R$ is much heavier. In Fig.~\ref{HAProduction} we show the variation of pair production cross-section and the associated production cross-section of the charged scalar $H^\pm_2$ with its mass at the LHC. We use \texttt{MadGraph5\_aMC@NLO v3.5.5\cite{Alwall:2014hca}} along with \texttt{SPheno-4.0.5\cite{Porod:2011nf}} to generate the production cross-sections at the collider experiments in Fig. \ref{HAProduction}.
The left panel of Fig. \ref{HAProduction} shows the variation of the pair production cross-section at the LHC experiment with the center of mass energies $\sqrt{s}=$ 13 TeV (solid blue line), 14 TeV (solid red line), 27 TeV (solid magenta line), and 100 TeV (solid black line). The bump and drop in the pair production cross-section can be attributed to $Z^\prime$ mediated channels of Drell-Yan production of the charge scalar $H^\pm_2$. To explicitly demonstrate it, we have shown the variation of pair production cross-section of the charge scalar $H^\pm_2$ with its mass while excluding $Z^\prime$ mediated channels with dashed lines, while including all channels with solid lines in the left panel of  Fig. \ref{HAProduction}. The corresponding Feynman diagrams are listed in Appendix \ref{production_channel}. Similarly, the right panel of Fig.~\ref{HAProduction} shows the cross section for the associated production of $H_2^\pm$  with the dark matter, which is comparable to that of charged Higgs pair production. This behavior is expected, since the dark matter particle and the charged Higgs are nearly degenerate in mass.
 
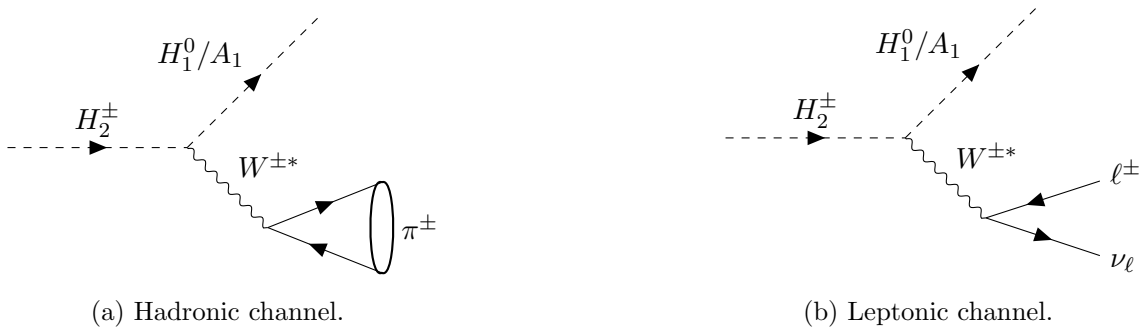
\begin{figure}[h!]
    \centering
    \begin{subfigure}{0.45\textwidth}
        \centering
        \begin{tikzpicture}
          \begin{feynman}
            \vertex (a) {};
            \vertex [right=2.5cm of a] (b);
            \vertex [above right=2.5cm of b] (c) {};
            \vertex [below right=1.5cm of b] (w);
            \vertex [right=1.5cm of w, yshift=0.6cm] (f1);
            \vertex [right=1.5cm of w, yshift=-0.6cm] (f2);

            \diagram* {
              (a) -- [charged scalar, edge label=\(H_2^\pm\)] (b),
              (b) -- [charged scalar, edge label=\(H_1^0/A_1\)] (c),
              (b) -- [boson, edge label=\(W^{\pm*}\)] (w),
              (w) -- [fermion] (f1),
              (w) -- [anti fermion] (f2),
            };

            \draw[thick] (f1) .. controls +(0.2,0.1) and +(0.2,-0.1) .. (f2)
                         .. controls +(-0.2,-0.1) and +(-0.2,0.1) .. (f1);
            \node at ($(f1)!0.5!(f2)+(.5,0)$) {\(\pi^\pm\)};
          \end{feynman}
        \end{tikzpicture}
        \caption{
        Hadronic channel.}
    \end{subfigure}%
    \hfill
    \begin{subfigure}{0.45\textwidth}
        \centering
        \begin{tikzpicture}
          \begin{feynman}
            \vertex (a) {};
            \vertex [right=2.5cm of a] (b);
            \vertex [above right=2.5cm of b] (c) {};
            \vertex [below right=1.5cm of b] (w);
            \vertex [right=1.5cm of w, yshift=0.6cm] (f1){$\ell^\pm$};
            \vertex [right=1.5cm of w, yshift=-0.6cm] (f2) {$\nu_\ell$};

            \diagram* {
              (a) -- [charged scalar, edge label=\(H_2^\pm\)] (b),
              (b) -- [charged scalar, edge label=\(H_1^0/A_1\)] (c),
              (b) -- [boson, edge label=\(W^{\pm*}\)] (w),
              (w) -- [anti fermion] (f1),
              (w) -- [fermion] (f2),
            };
          \end{feynman}
        \end{tikzpicture}
        \caption{
        Leptonic channel.}
    \end{subfigure}
    
    \caption{Feynman diagrams of $H_2^\pm$ decays with mass splitting between the charged and neutral scalars, $\Delta M <1$ GeV. The $R$-parity odd fermions, $n$ and $d'$ are taken to be heavier than $H_2^\pm$.}
    \label{fig:w-decays}
\end{figure}

The charged scalar $H^\pm_2$ further decays into the dark sector particles depending on the mass spectrum of particles in the dark sector. Its fermionic and scalar decay channels are generically \( \Gamma(H^+_2 \rightarrow  n_i~l_j),~\Gamma(H^+_2 \rightarrow u_i \bar{d^\prime_j}),~{\rm and }~\Gamma(H^+_2 \rightarrow H_1^0/A_1  ~W^+_L).
\)
The purely gauge boson decay channels are not relevant in the parameter region explored in this study, since such decays necessarily involve a virtual or real $W_R$, which in our setup is assumed to be much heavier than $H_2^\pm$. Moreover, as discussed in the previous section, the scotino dark-matter scenario is not viable. We therefore consider the case where the scotinos $n_{e,\mu, \tau}$ and $d'$ are heavier than $H_2^\pm$, leaving the scalar dark-matter channel as the only viable possibility.
As our interest lies in LLP signatures, we focus on small mass splitting,  $\Delta M=m_{H_2^\pm}-m_{H_1^0}<1$ GeV. In this regime, the allowed decay channels are
\[
H_2^\pm\rightarrow H_1^0/A_1~W^*_L\to H_1^0/A_1~(q\bar q',~\ell \nu_\ell).
\]
For $\Delta M < 1$ GeV, the quarks produced from the off-shell $W^*_L$ do not hadronize into energetic jets. Instead, the $q\bar q'$ system hadronizes into light mesons, predominantly $\pi^\pm$, once the charged-pion threshold is crossed. Below the pion threshold, the decay proceeds exclusively through the leptonic channels. The Feynman diagrams relevant to the decays of the charged scalar $H^\pm_2$ are shown in Fig. \ref{fig:w-decays}. For details we refer to \cite{ Belyaev:2020wok,Cirelli:2005uq}. 
The decay width of the $\pi^\pm$ channel is computed with the help of an effective Lagrangian given by \cite{Belyaev:2020wok} 
\begin{equation}
  \mathcal{L}_{H_2^+(H_1^0/A_1)\pi^+}= \frac{ig_L ~g_{(HAW)}f_\pi}{4\sqrt{2}m_W^2}(p_{H_2^+}-p_{A_1}).p_
  {\pi^+},
\end{equation}
where $f_\pi$ = 130 MeV is the pion decay constant,  $g_{HAW}=g_L\rm cos~\xi$ (with $\xi$ given by $\tan\xi=\tfrac{k}{v_R}$) and $p_i$ are the momenta of the corresponding particles.
This leads to the decay width given by \cite{Belyaev:2020wok} 
\begin{eqnarray} 
\Gamma(H_2^\pm\rightarrow H_1^0/A_1~\pi^\pm)&=& \frac{g_{HAW}^2~ g_L^2~ f_\pi^2}{64 \pi ~m_W^4}\Delta M^2\sqrt{\Delta M^2-m_{\pi^\pm}^2}
\end{eqnarray}

 \par
In this study, we shall exclusively consider the hadronic channel, $H_2^\pm\to H_1^0/A_1~\pi^\pm$ with the mass splitting within the range of $0.2<\Delta M<1$ GeV.

\subsection{Numerical Studies}\label{pheno1}
\begin{figure}[H]
\centering
\includegraphics[width=.49\textwidth, scale=0.85]{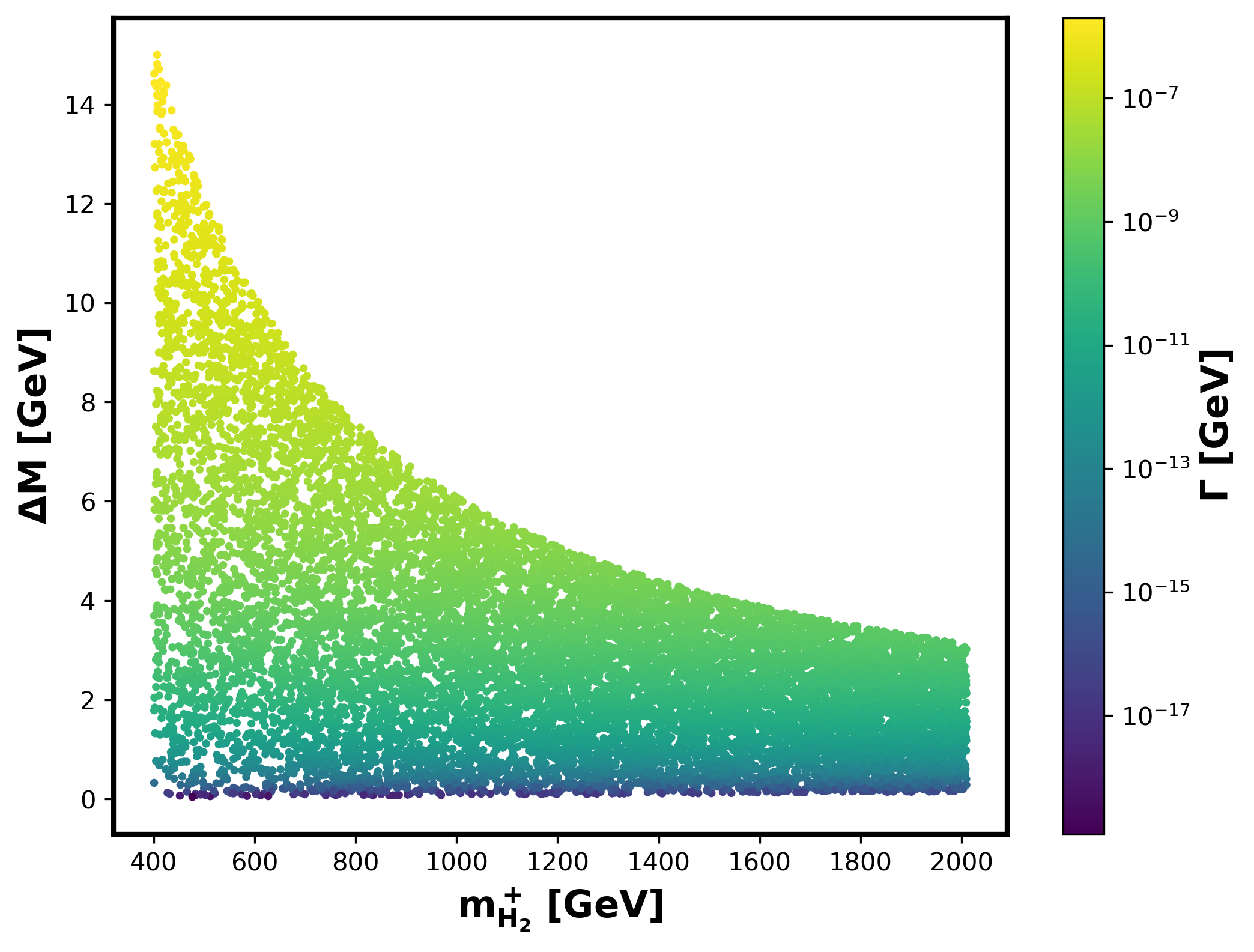}
\caption{\label{GammaScan}  Variation of the total decay width $\Gamma$ of $H^\pm_2$ as a function of its mass $m_{H^\pm_2}$ and the mass splitting $0.2<\Delta M=(m_{H^\pm_2}-m_{\rm DM})<15$ GeV. 
For this scan, we set $\lambda_2$ in the range [-$10^{-5}$, -0.1] so as to get the $\Delta M$ in the desired range.
Other parameters are taken to be: $\alpha_{1,2,3}=0.1,~\lambda_3=0.1, ~ \text{tan}\beta=10, {~\rm and} ~ v_R=20$ TeV. }
\label{fig:DeltaM}
\end{figure}

\begin{figure}[H]
\centering
\includegraphics[width=.49\textwidth, scale =0.8]{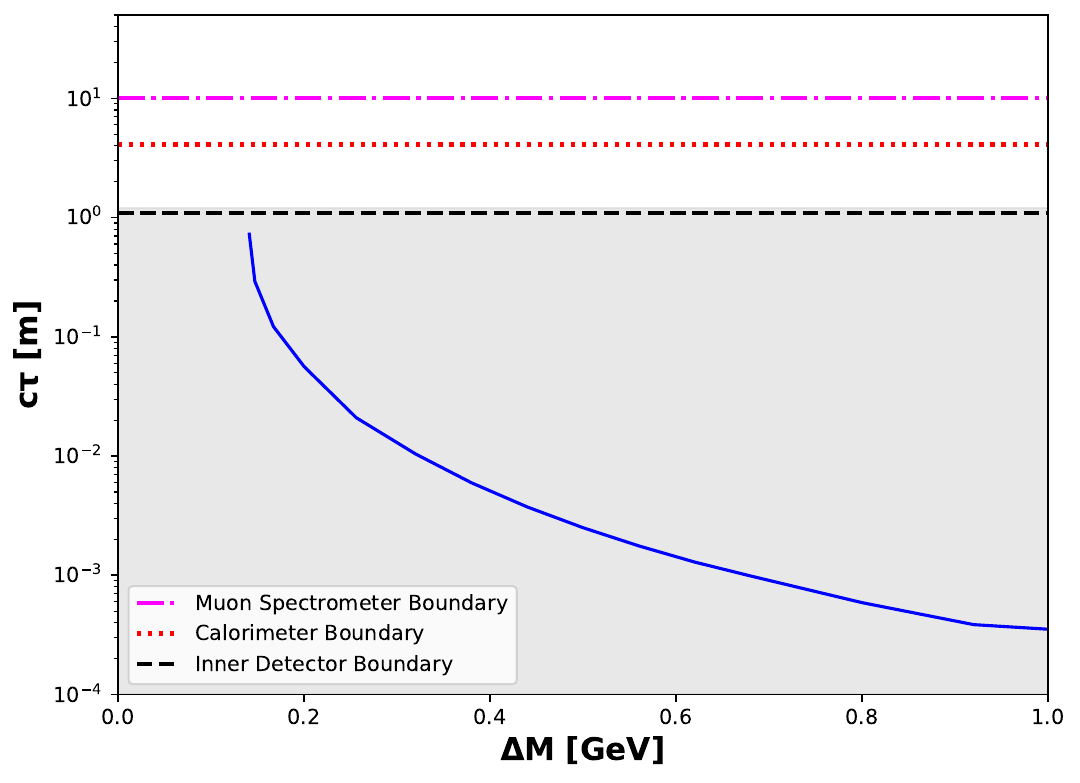}
\caption{\label{HAproper_ctau} The proper decay length ($c\tau$) of the charged scalar $H^\pm_2$ as a function of the mass splitting $\Delta M$ for  $ m_{H_2^\pm}=1$ TeV. Other parameters are same as in Fig.~\ref{GammaScan}. The gray shaded region corresponds to the inner tracker of ATLAS detector \cite{Aad:1129811}.}
\end{figure}

In order to characterize the long-lived behavior of the charged Higgs boson $H_2^\pm$, we begin by exploring the dependence of its decay width on the mass splitting between $H_2^\pm$ and the dark sector states. A preliminary scan over the relevant parameters is performed to identify the range of mass splittings that lead to sufficiently suppressed decay widths compatible with macroscopic lifetimes. This analysis indicates that long-lived charged Higgs signatures typically arise for mass splittings below the GeV scale. 
We  consider the parameter values as $\lambda_3=0.1$, ${\rm tan}~\beta=10$, $\alpha_{1,2,3}=0.1,~v_R=20$ TeV and varied the parameters $\lambda_2$ and $m_{H_2^\pm}$ in the range [$-10^{-5},~-0.1$] and [400 GeV, 2000 GeV], respectively. As discussed in Section~\ref{DM}, the range of $\lambda_2$ is chosen to limit the mass splitting $\Delta M= m_{H_2^\pm}-m_{H_1^0/A_1}\lesssim 10$ GeV. Using \texttt{CalcHEP-3.9.2} and  \texttt{SPheno-4.0.5} the parameter space is scanned for decay width of $H_2^\pm$. 
Fig. \ref{GammaScan} shows the variation of the total decay width of $H^\pm_2$ as a function of its mass $m_{H^\pm_2}$ and mass splitting $\Delta M$.
Notice that only $\Delta M\lesssim 1$ GeV yield a total decay width  $\Gamma_{H^\pm_2}\lesssim 10^{-13}$ GeV required to distinguish $H_2^\pm$ as a long-lived particles at the LHC experiments. To further demonstrate this, we consider a specific benchmark point with $m_{H^\pm_2}= 1$ TeV and compute $\Gamma_{H_2^\pm}$ with varied mass splitting, $\Delta M$. 
Fig. \ref{HAproper_ctau} shows the proper decay length $c\tau$ of the charged scalar $H^\pm_2$ as a function of the mass splitting $\Delta M$ as shown with a solid blue line. We have also marked the inner detector boundary, the calorimeter boundary, and the muon spectrometer boundary of the ATLAS detector \cite{Aad:1129811} with  the dashed black line, the dotted red line, and the dot-dashed magenta line, respectively. With such small mass splitting, the SM particles originating from the decay of $H^\pm_2$ are too soft to be detected, resulting in a disappearing track-like signature \cite{Belyaev:2020wok}. Such disappearing track signatures are studied by both the CMS~\cite{CMS:2018rea}\cite{CMS:2020atg} and the ATLAS~\cite{ATLAS:2017oal}\cite{ATLAS:2022rme} experiments at the LHC.

\begin{figure}[H]
\centering
\includegraphics[width=.49\textwidth]{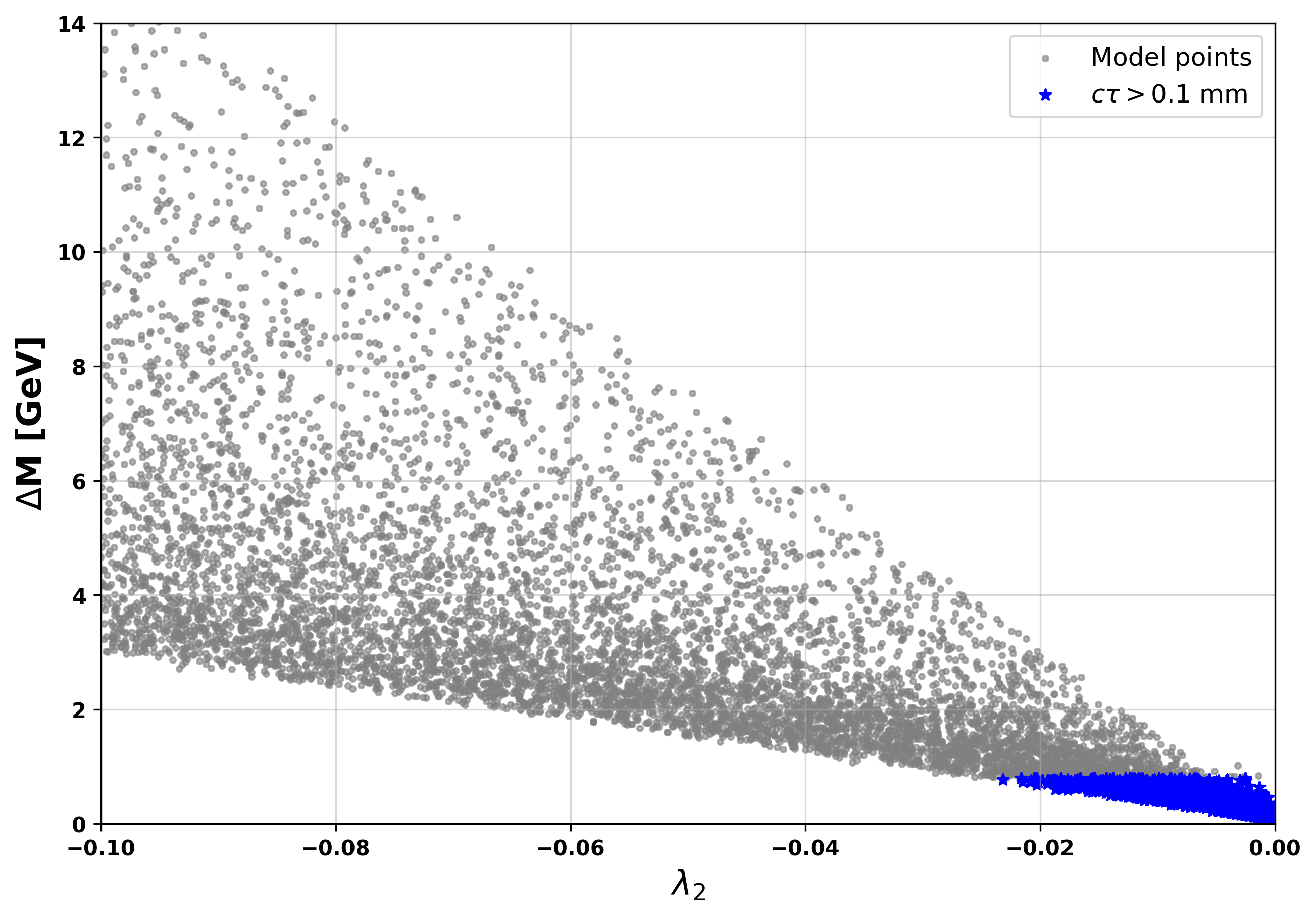}
\includegraphics[width=.49\textwidth]{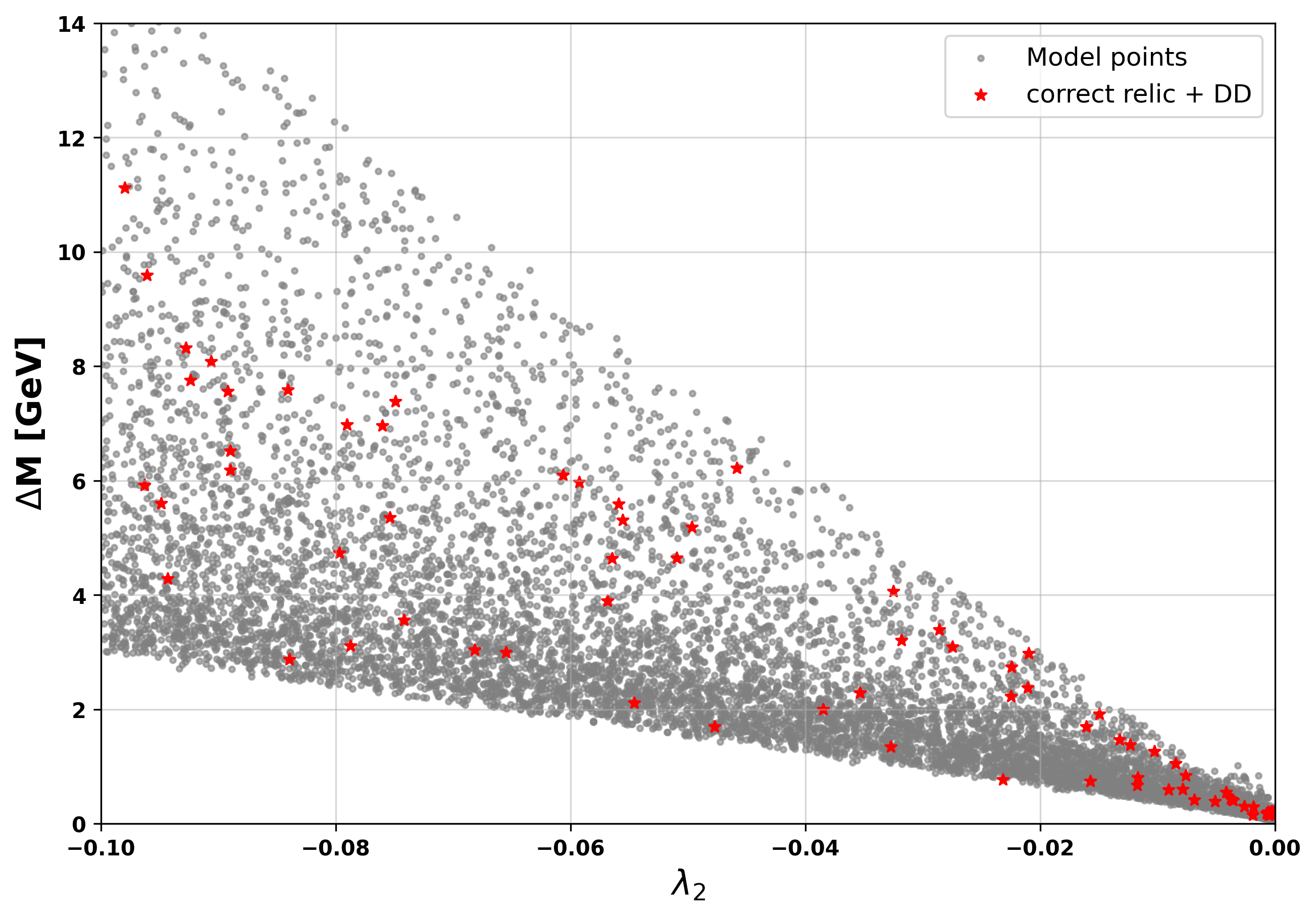}
\caption{\label{HADM_lam2_new} The scatter plot in $\lambda_2- \Delta M$ plane ($\Delta M = m_{H_2^\pm}-m_{DM}$) obtained from parameter scan as shown in  Fig.~\ref{fig:DeltaM}. \textit{(Left panel)}: the blue point correspond to a proper decay length of $H_2^\pm$, ($c\tau >0.1$ mm). \textit{(Right panel)}: the red points satisfy the observed relic density with $0.119<\Omega h^2 < 0.121$ and the XENONnT direct detection limit. }
\end{figure}

We aim to analyze a few benchmark points with sizable $H_2^\pm$ pair production cross section, and at the same time leading to disappearing track signatures, as described above. For these points to be compatible with the dark matter constraints, we select out the points from those studied in scalar dark matter case in Section~\ref{DM}. Plotting all the points of Fig.~\ref{HArelic_mA_new} in the $\lambda_2$-$\Delta M$ plane, we select out those with proper decay length $c\tau>0.1$mm, as presented in Fig.~\ref{HADM_lam2_new}. In the left panel of Fig.~\ref{HADM_lam2_new}, the blue points in the $\lambda_2$-$\Delta M$ plane correspond to the regions of the  parameter space where the $H^\pm_2$ behaves like a long-lived particle (LLP) giving rise to disappearing track-like signature with proper decay length $c\tau> 0.1$ mm. It is evident that such LLP's require small mass splitting $\Delta M<1$ GeV. In the right panel of Fig.\ref{HADM_lam2_new} the red star-shaped points indicate parameter regions that simultaneously satisfy the observed darkmatter relic density  and the XENONnT direct detection limits. 
For our LLP study, we have selected nine benchmark points  satisfying the above requirements, as listed in Table \ref{HAtable4}. These benchmark points are grouped into three categories corresponding to  $m_{H_2^\pm}=1,~~1.2, ~1.4$ TeV.  For each set, three benchmark points with different $m_{DM}$ are chosen to gauge the effect of their mass splitting, $\Delta M$. The corresponding masses of other BSM states  are listed in  Table~\ref{HAtable_mass_subset}.  

\begin{table}[h!]
\centering
\begin{tabular}{c|c|c|c|c|c|c|c|c|c}
\hline\hline
  {\textbf{BPs}} & $m_{H_2^+}$& $m_{{H_1^0}}=m_{A_{1}}$&$\lambda_2$& $\lambda_3$& tan$\beta$& $v_R$&$\Gamma_{\rm tot}$ & $\Omega h^2$ &$c\tau$ \\
  &(GeV) &(GeV) & & & &(GeV) &(GeV) & &(cm)\\
  \hline \hline

    \multirow{3}{*}{{\bf{BP1}}}  &\multirow{3}{*}{1000}  
   &(a) $999.80 $& -0.0019 & $0.022$ &22.8 & 19659.9 &$4.12\times 10^{-15} $&0.120&4.78  \\ 
    &&(b) $999.59$& -0.0056&0.021&22.2& 20534.4 &$4.67\times 10^{-14}$ &0.119&0.42  \\
     &&(c) $999.41$& -0.0091&0.017&34.5& 26529.3 &$1.44\times 10^{-13}$ &0.119&0.14  \\\hline
 
  \multirow{3}{*}{{\bf{BP2}}}  &\multirow{3}{*}{1200}  
   &(a) $1199.81 $& -0.00037 & $0.037$ &31.1 & 14523.0 &$ 3.34\times 10^{-15} $&0.120 &5.89 \\  
   &&(b) $ 1199.70$ & -0.0010 &0.042& 20.8& 12210.8&$1.72\times 10^{-14}$&0.120 &1.18  \\
    &&(c) $1199.58$& -0.0069 &0.020&16.0& 23065.1 & $5.04\times 10^{-14}$ &0.120&0.39  \\ \hline

    \multirow{3}{*}{{\bf{BP3}}}  &\multirow{3}{*}{1400}  
   &(a) $1399.80 $& -0.00010 & $0.041$ &10.5 & 14650.4 &$ 4.12\times 10^{-15} $&0.120 &4.78 \\ 
   &&(b) $ 1399.69$ & -0.0026 &0.039& 26.2& 14998.8&$1.91\times 10^{-14}$&0.119 &1.02  \\
    &&(c) $1399.59$& -0.0054 &0.038&16.8& 15698.4 & $4.67\times 10^{-14}$ &0.119&0.42  \\ \hline\hline
\end{tabular}
\caption{\label{HAtable4} Benchmark points (BPs) for the disappearing track-like signature from the decay of $H^\pm_2$. All points are compatible with the observed relic density and the direct detection limit.}
\label{tb:BPs}
\end{table}

\begin{figure}[H]
 \includegraphics[width=.33\textwidth]{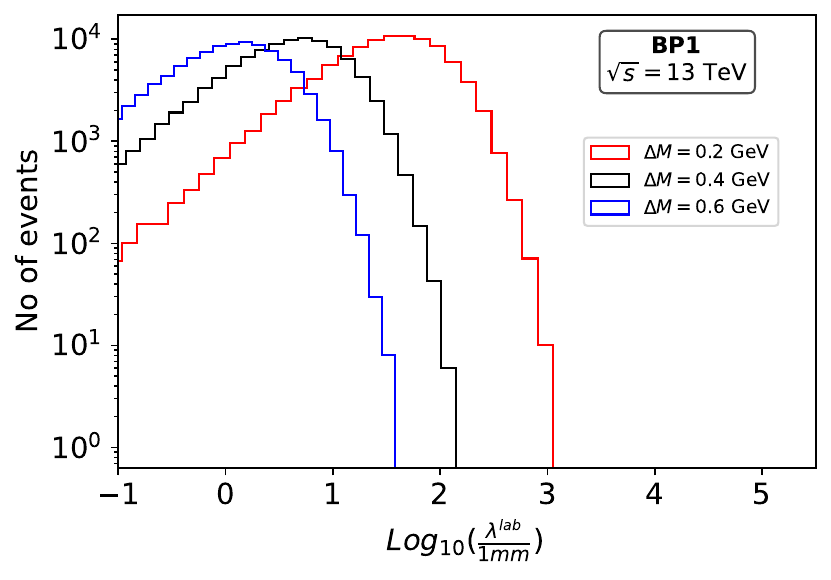}
 \includegraphics[width=.33\textwidth]{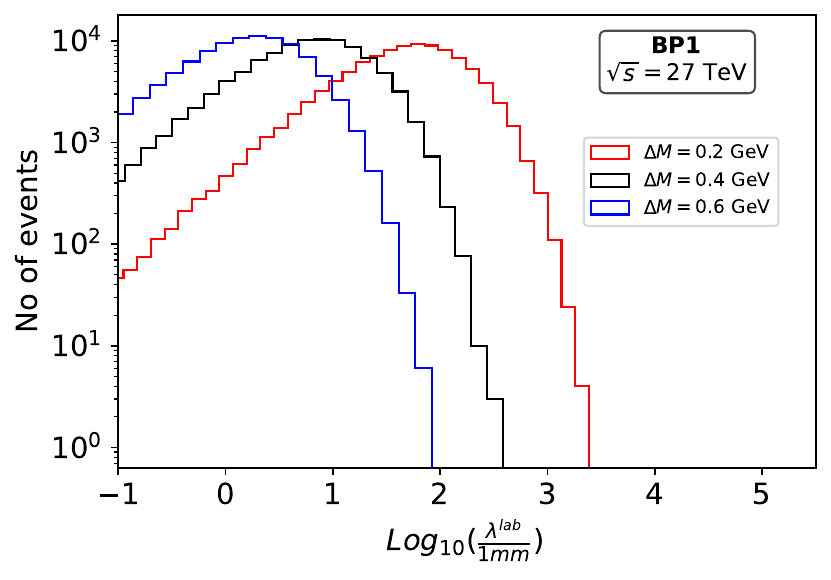}
 \includegraphics[width=.33\textwidth]{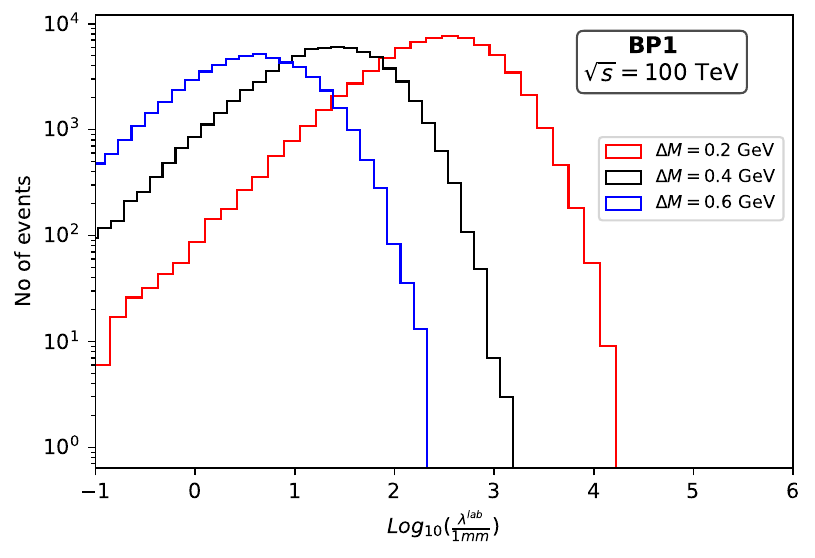}

 \includegraphics[width=.33\textwidth]{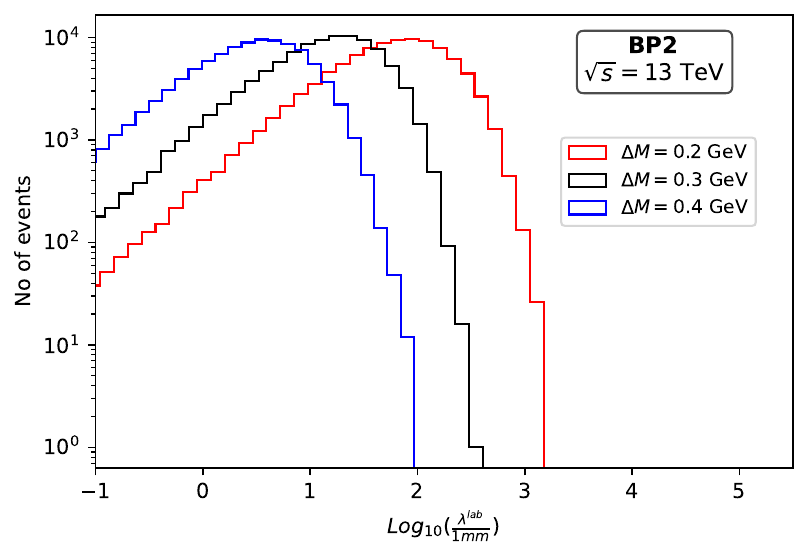}
 \includegraphics[width=.33\textwidth]{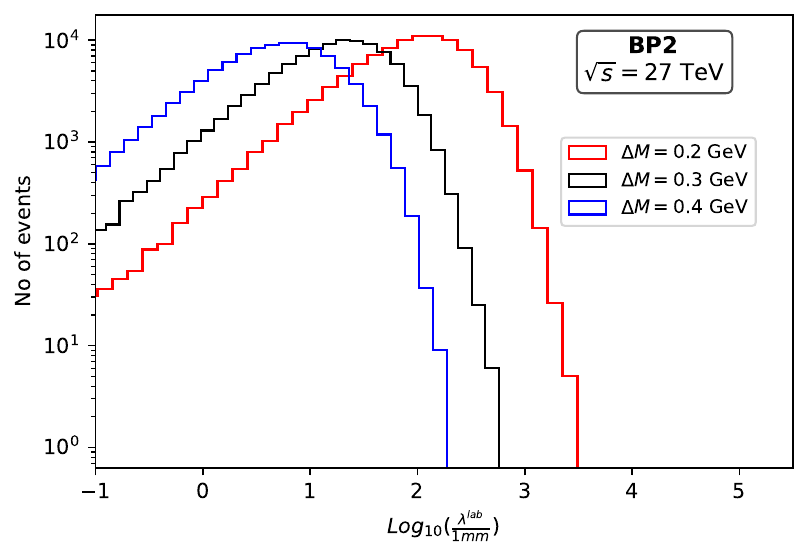}
 \includegraphics[width=.33\textwidth]{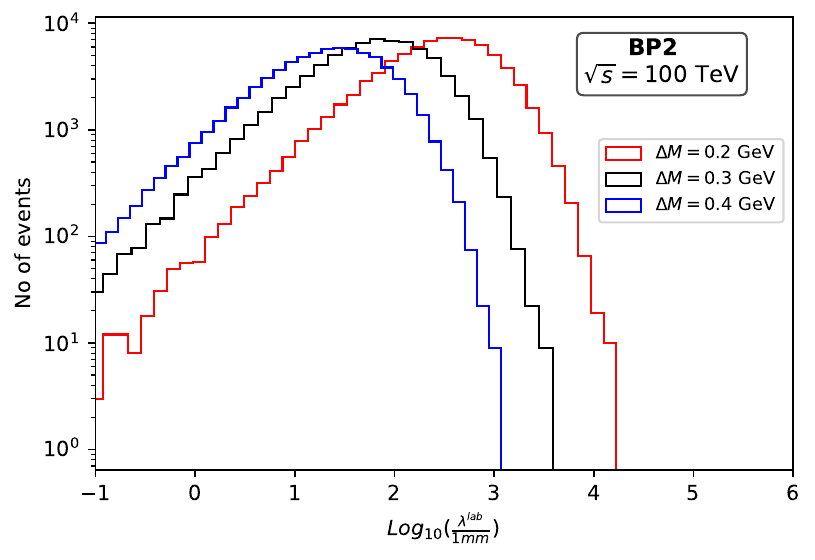}
 
 \caption{\label{HAboosted_length} Distribution of 50,000 events with the boosted decay lengths (log$_{10}\lambda$) for the mass of charged scalar $m_{H_2^+}$= $1000$ GeV (top), and $1200$ GeV (bottom), respectively at the $\sqrt{s}=$13 TeV (left),  27 TeV (middle) and 100 TeV (right) LHC with three different values of $\Delta M$.}
\end{figure}

\begin{table}[h!]
\centering
\renewcommand{\arraystretch}{1.2}
\begin{tabular}{c|c|c|c|c|c|c|c|c}
\hline \hline
\multicolumn{9}{c}{\large{Particle Mass (GeV)}} \\ \hline
\textbf{BPs} & $m_{H^0_2}$ & $m_{{H^0_3}}$ & $m_{{A_{2}}}$ & $m_{H_1^\pm}$ & $m_{n_e}$ & $m_{n_\mu,n_\tau}$ & $m_{W_R}$ & $m_{Z'}$ \\ 
  \hline\hline
  
   \multirow{3}{*}{{\bf{BP1}}} 
   &(a) $4122.8$ & $22777.1$ & $22777.1$ & $22777.1$ & $1013.5$ & $1018.6$ & $6389.2$ & $7634.2$ \\  
   &(b) $4215.8$ & $22218.9$ & $22218.9$ & $22218.9$ & $1010.5$ & $1016.2$ & $6674.1$ & $7973.5$ \\
   &(c) $4920.4$ & $34552.9$ & $34552.9$ & $34552.9$ & $1007.1$ & $1014.7$ & $8622.4$ & $10301.4$ \\ \hline
   \multirow{3}{*}{{\bf{BP2}}} 
   &(a) $3946.6$ & $37273.2$ & $37273.2$ & $37273.2$ & $1220.5$ & $1236.6$ & $4720.6$ & $5639.6$ \\  
   &(b) $3565.5$ & $24923.6$ & $24923.6$ & 24923.5 & $1212.91$ & $1214.6$ & $3969.3$ & $4741.4$ \\
   &(c) $4582.6$ & $19260.6$ & $19260.6$ & $19260.6$ & $1209.3$ & $1213.5$ & $7496.5$ & $8956.3$ \\ \hline

   \multirow{3}{*}{{\bf{BP3}}} 
   &(a) $4199.0$ & $14694.8$ & $14694.8$ & $14694.7$ & $1420.5$ & $1433.2$ & $4762$ & $5689.1$ \\  
   &(b) $4219.5$ & $36631.8$ & $36631.8$ & 36631.8 & $1418.5$ & $1430.2$ & $4875.1$ & $5824.9$ \\
   &(c) $4314.1$ & $23488.8$ & $23488.8$ & $23488.8$ & $1416.5$ & $1439.2$ & $5102.6$ & $6095.9$ \\
   \hline \hline
\end{tabular}
\caption{\label{HAtable_mass_subset} The masses of different particles corresponding to benchmark points in Table \ref{HAtable4}.}
\end{table}
Since $H_2^\pm$ is produced with a broad momentum distribution at the LHC, its Lorentz boost varies event by event and has a direct impact on the decay length in the laboratory frame. Consequently, the decay length observed in the detector can be significantly different from the corresponding proper decay length listed in Table~\ref{HAtable4}.  For all the BP's, we therefore evaluate the boosted decay length,$\lambda^{\rm lab} = \beta \gamma c \tau$,  where $\beta=\frac{v}{c}$ and $\gamma=\frac{1}{\sqrt{1-\beta^2}}$ is the Lorentz factor, $v$ being the speed of $H_2^\pm$ in the lab frame. We consider the LHC at  $\sqrt{s}=$ 13 TeV and its high energy version at $\sqrt{s}=27, ~100$ TeV.
For the numerical analysis, we have used \texttt{CalcHEP-3.9.2}~\cite{Belyaev:2012qa} together with \texttt{SPheno-4.0.5}~\cite{Porod:2011nf} to simulate the pair production of the charged scalar $H^\pm_2$ at the LHC. We employ the \texttt{NNPDF23\_lo\_as\_0130\_qed} parton distribution function~\cite{NNPDF:2014otw}, with the QCD scale fixed at twice the mass of the charged scalar, $\mu=2m_{H^\pm_2}$. The boosted decay lengths are subsequently obtained using  \texttt{Pythia-8.3}~\cite{Bierlich:2022pfr}.
Fig. \ref{HAboosted_length} displays the distribution of boosted decay length ($\lambda^{\rm lab}$) of 50000 randomly simulated events corresponding to BP1 (top) and BP2 (bottom), at the center of mass energy $\sqrt{s}$ = 13 TeV, 27 TeV and 100 TeV. The results for BP3 exhibit a similar behavior and are therefore not shown explicitly. As expected, the impact of the larger momentum (boost) is clearly visible in the $\sqrt{s}=100$ TeV case compared to the $\sqrt{s}=13,~27$ TeV case.

\begin{figure}[H]

\subcaptionbox{}{\includegraphics[width=.3\textwidth]{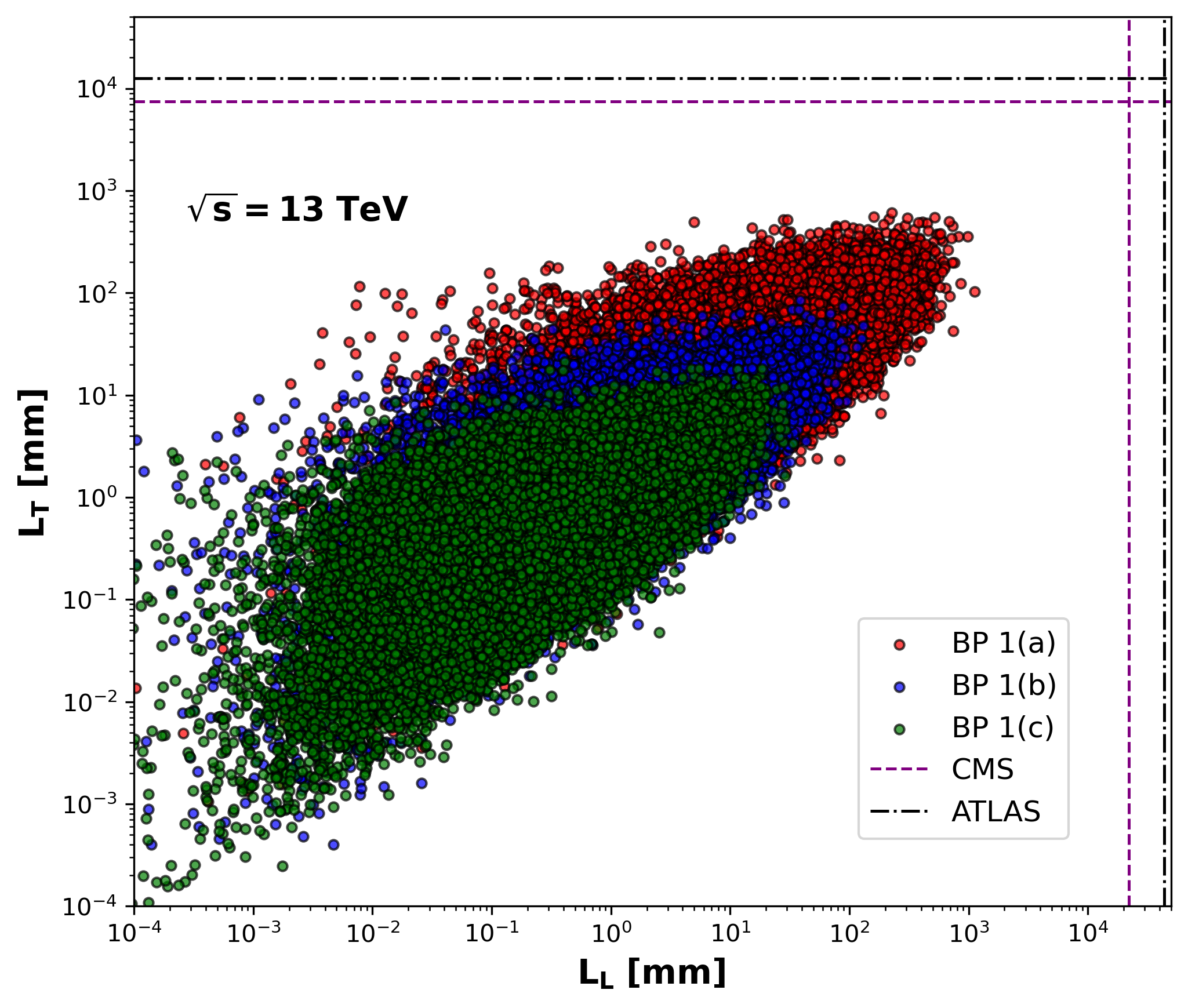}} \qquad
\subcaptionbox{}{\includegraphics[width=.3\textwidth]{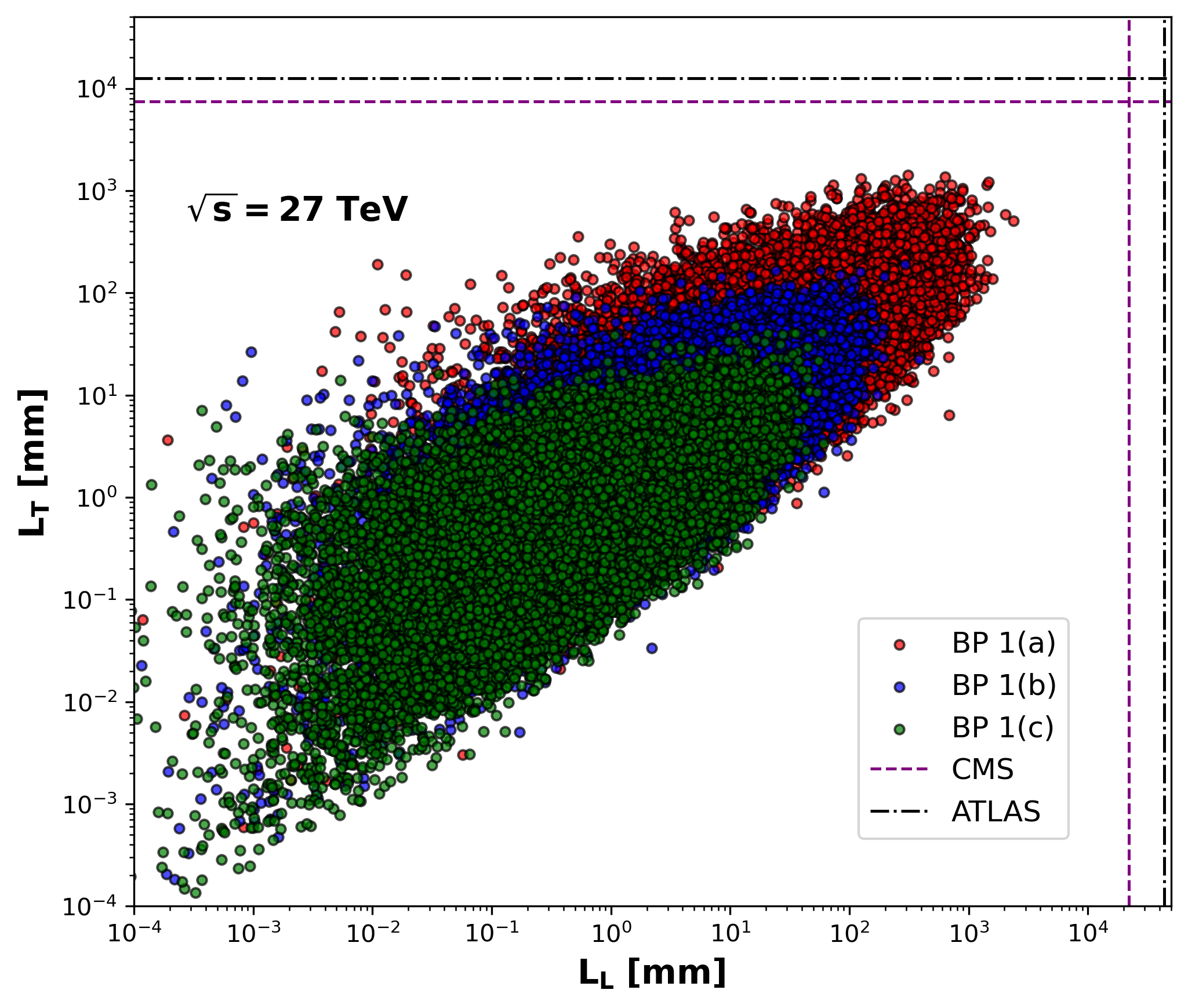}} \qquad
\subcaptionbox{}{\includegraphics[width=.3\textwidth]{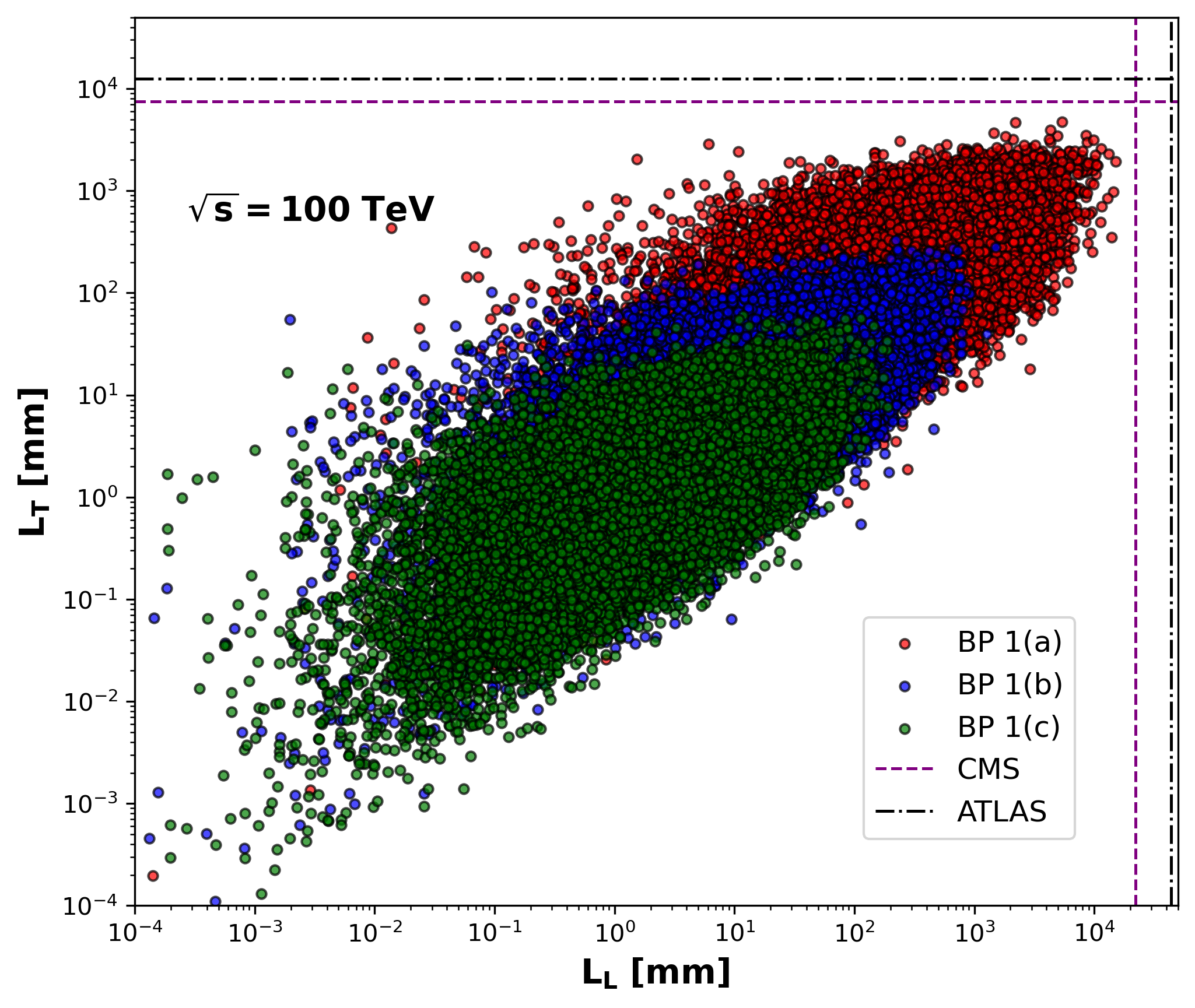}} \qquad
\subcaptionbox{}{\includegraphics[width=.3\textwidth]{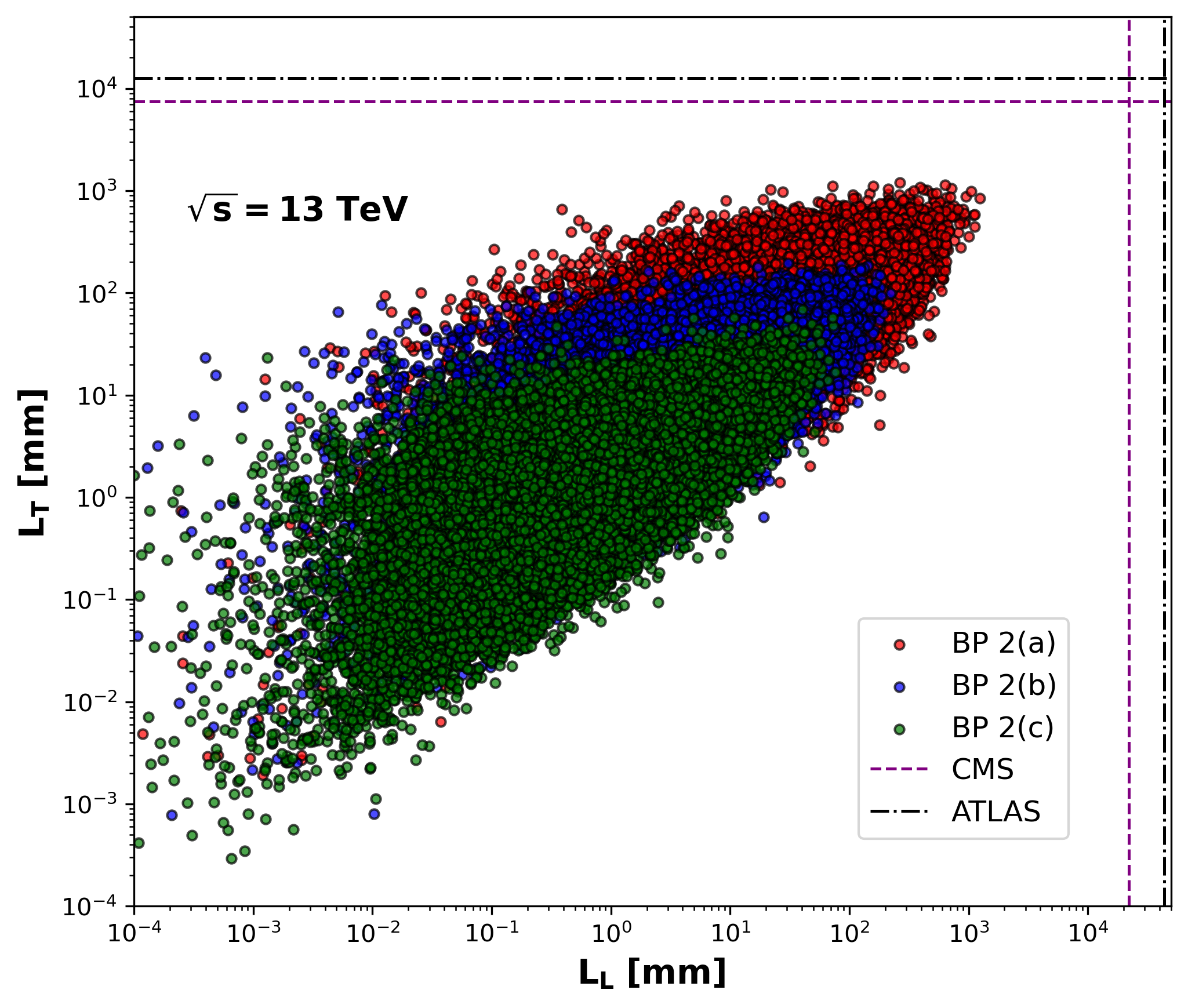}} \qquad
\subcaptionbox{}{\includegraphics[width=.3\textwidth]{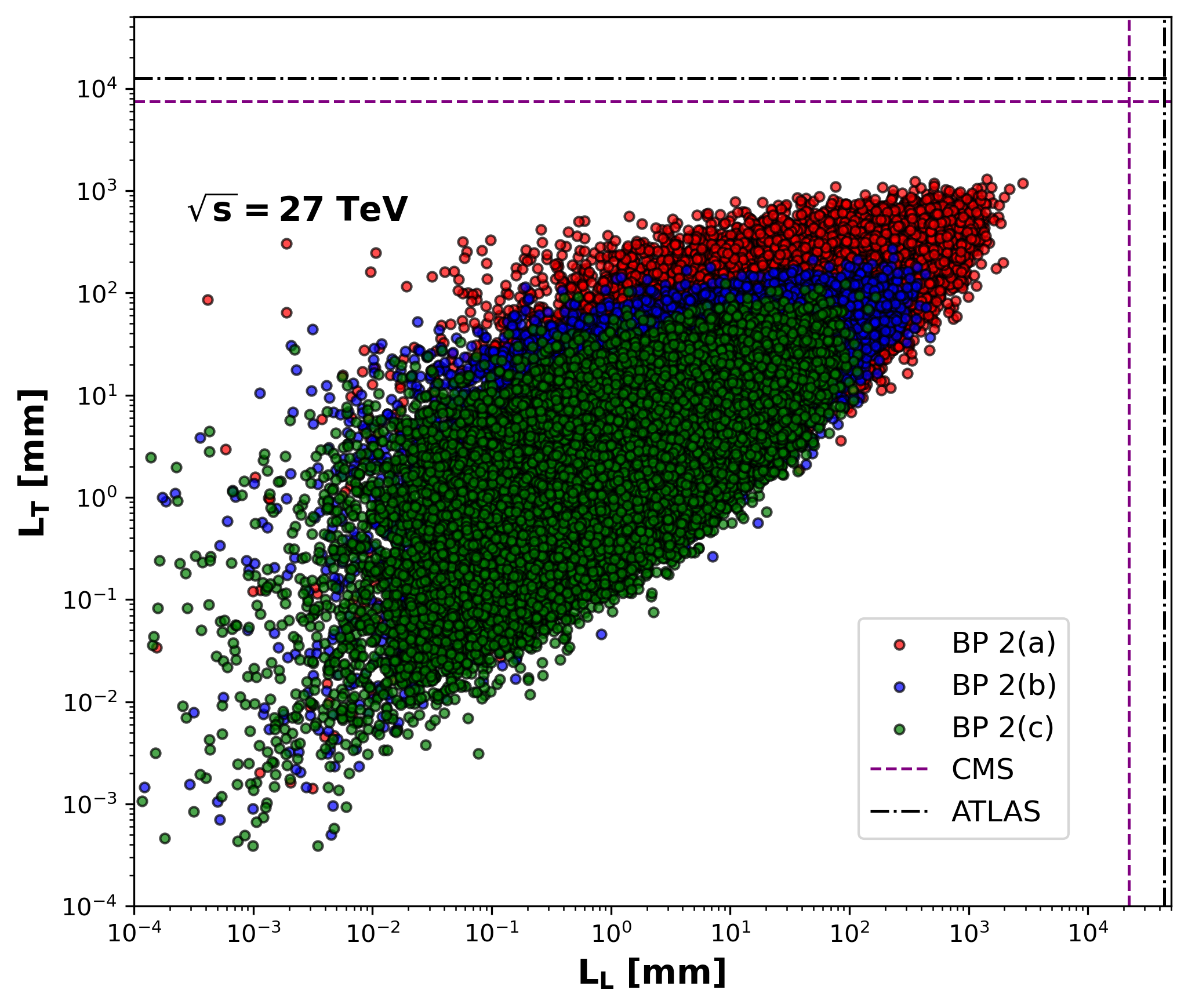}} \qquad
\subcaptionbox{}{\includegraphics[width=.3\textwidth]{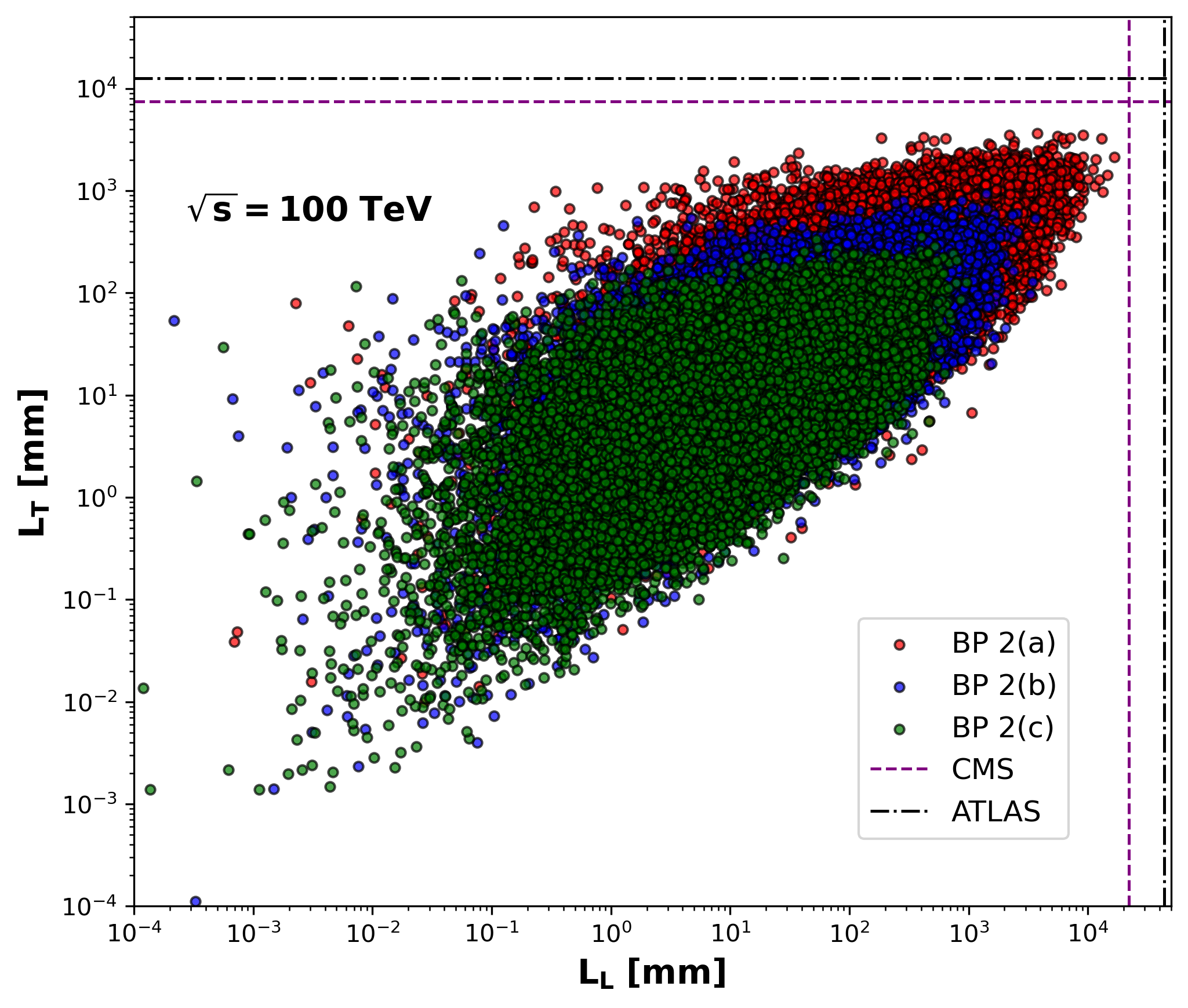}} \qquad

\caption{\label{HAtransvers_long_length_BP} Correlation plot between the longitudinal decay length, $(L_L= \frac{p_z}{m}c\tau$ ) and the transverse decay length, $(L_T= \frac{p_T} {m}c\tau$) for the selected BPs in the centre of mass energy $\sqrt{s}=$ 13, 27 and 100 TeV respectively, where $p_T=\sqrt{p_x^2+p_y^2}$ is the transverse momentum. The upper panel is for BP1 and lower is for BP2. The magenta dashed, and the black dash-dotted lines correspond to the longitudinal and transverse coverages of the CMS and the ATLAS detectors,   respectively. }
\end{figure}

In Fig. \ref{HAtransvers_long_length_BP}, we show the distribution of the events with longitudinal and transverse displaced (boosted) lengths for BP1 and BP2. The CMS and the ATLAS detectors have approximate transverse decay length coverages of about 7.5 m and 12.5 m, respectively, with corresponding longitudinal ranges of about 22 m and 44 m~\cite{_2007,ATLAS:TDR1999}. In the figure, the dashed and dash-dotted lines symbolize the upper limits of the CMS and the ATLAS detectors, respectively. 
Such a long-lived $H_2^\pm$, decaying inside the detector at a measurable distance from the interaction point, can give rise to disappearing-track signatures. Both the ATLAS and CMS collaborations have performed dedicated searches for such exotic signals. In the next section, we examine the implications of these searches for the $H_2^\pm$ in the ALRM.

\subsection{Disappearing track analysis} \label{DTA}

As discussed in the previous section, in regions of parameter space with a compressed mass spectrum, the charged scalar \(H_2^\pm\) can become long-lived on detector length scales. In such cases, its decay products are typically too soft to be reconstructed, causing the charged particle track to terminate abruptly within the inner tracking volume or in the calorimeters. This results in a characteristic disappearing track signature, which provides a powerful and relatively background-free probe of long-lived charged particles at the LHC.
The dominant backgrounds in disappearing track searches are largely instrumental in nature and therefore cannot be reliably modeled using Monte Carlo simulations. Instead, they are estimated directly from data control samples. The disappearing track signature has been extensively investigated by the CMS~\cite{CMS:2018rea}\cite{CMS:2020atg} and the ATLAS~\cite{ATLAS:2017oal}\cite{ATLAS:2022rme} collaborations at the LHC, primarily in the context of long-lived charginos decaying into a neutralino and a charged pion. Both experiments provide lifetime-dependent exclusion limits for wino- and higgsino-like scenarios.
When comparing our benchmark points with the published experimental limits (Fig.~10), we carefully ensure consistency with the assumptions adopted in the corresponding analyses. In particular, we match the production mechanism, the lifetime definition, and the key analysis selections, such as the MET+jet trigger requirements, track hit criteria, and calorimeter vetoes, which have a significant impact on the signal acceptance and, consequently, on the derived exclusion limits. We shall describe some details of the procedure adopted in the following section.

\subsubsection{Observability of the disappearing tracks}\label{subsec:observability}

As already mentioned, while there is no irreducible Standard Model background at the parton level for this signal, the disappearing track signature can be mimicked by detector-related effects, such as track reconstruction inefficiencies, interactions with detector material, or mismeasured tracks. These instrumental backgrounds dominate the experimental searches and must be estimated using data-driven methods rather than Monte Carlo simulations. The observability of the signal will further depend on the efficiency with which the track is reconstructed.  The two parameters relevant to the reconstructed events are the event acceptance efficiency and the tracklet efficiency\cite{CMS:2020atg, ATLAS:2022rme} \footnote{In this work, we focus exclusively on the ATLAS disappearing-track search~\cite{ATLAS:2022rme}, as the corresponding CMS~\cite{CMS:2020atg} and ATLAS analyses yield comparable sensitivities for the class of long-lived charged-particle scenarios considered here.}. For the purpose of this discussion, a tracklet is basically a short track inside the detector. In our analysis, we closely follow the methodology developed and used in Ref.~\cite{Belyaev:2020wok}, which is available in the LLP repository\cite{LLPRecasting:DisappearingTracks}. 
This reinterpretation strategy allows us to compute the event acceptance and tracklet reconstruction efficiency for disappearing-track signatures, and hence the visible cross section, defined as the product of the total production cross section and the overall efficiency. In this work, we compare our results exclusively with the latest ATLAS disappearing-track analysis based on an integrated luminosity of 136 fb$^{-1}$ at the centre of mass energy of 13 TeV~\cite{ATLAS:2022rme}, which provides the most stringent experimental constraints to date.

For the numerical analysis, we use \texttt{CalcHEP-3.9.2}~\cite{Belyaev:2012qa} with \texttt{SPheno-4.0.5}~\cite{Porod:2011nf} for parton-level event generation. Parton showering and hadronization are performed with \texttt{Pythia-8.3}~\cite{Bierlich:2022pfr}, followed by a detector-level simulation using \texttt{Delphes-3.5.0}~\cite{deFavereau:2013fsa} with the default ATLAS detector card. In addition, we employ the \texttt{NNPDF23\_lo\_as\_0130\_qed} parton distribution function set~\cite{NNPDF:2014otw}, with the QCD renormalization and factorization scales fixed to twice the charged scalar mass.
To study the disappearing track-like signals at the LHC, the relevant signal processes for the current framework are $p~p\rightarrow H^+_2 H^-_2$ and $p~p\rightarrow H^+_2 ~A_1/H^0_1$. 
The total production cross section for the combined process at the 13 TeV LHC, the HL-LHC at 14 TeV and its high energy versions (HE-LHC) at 27 TeV and 100 TeV centre of mass energies are given in Table~\ref{tab:bps_sigma}. 
\begin{table}[h!]
\centering
\begin{tabular}{l|c|c|c|c}
\hline\hline
&\multicolumn{4}{|c}{Total production cross section (fb)} \\
\cline{2-5}
\textbf{BPs}&$\sqrt{s}=13$ TeV& 14 TeV& 27 TeV&100 TeV\\
\hline\hline

& (a)  0.0328& 0.0445& 0.401  & 7.74\\
\textbf{BP1}& (b)   0.0325 & 0.0442& 0.390 & 7.31 \\
& (c)  0.0329 &0.0445& 0.372 & 5.81 \\
\hline
& (a)  0.0123& 0.0188& 0.389  & 11.13 \\
\textbf{BP2}& (b)  0.0239 & 0.0378& 0.732 & 17.59 \\
& (c)  0.0097 & 0.0138& 0.159 & 3.71 \\
\hline
& (a)  0.0052 & 0.0087& 0.270  & 9.02 \\
\textbf{BP3}& (b)  0.0047 & 0.0079& 0.240 & 8.47 \\
& (c)  0.0040 &0.0065& 0.199 & 7.23 \\
\hline\hline
\end{tabular}
\caption{Total production  cross sections, $\sigma_{\rm prod} (pp\to H^+_2H_2^-,H^\pm_2~H_1^0/A_1)$ (fb), at different colliders for the selected benchmark points in Table~\ref{tb:BPs}.}
\label{tab:bps_sigma}
\end{table}
To tag such events at the detector, we use initial-state radiation (ISR) induced mono-jet, along with large missing transverse momentum. The inclusion of this extra ISR jet is to lower the MET as much as possible so that we can have a significant number of signal events. Following the simplified methodology as in Ref. \cite{Belyaev:2020wok}, the event pre-selection demands at least an isolated tracklet, a high $p_T$ jet, and MET.  
After the event pre-selection stage, two further steps are applied: the event selection stage and the tracklet selection stage. The former isolates the signal from the Standard Model backgrounds, while the latter selects high-quality tracklets. The details of the event selection and tracklet selection criteria can be found in Refs. \cite{Belyaev:2020wok, ATLAS:2022rme}, with the main features summarized in Appendix \ref{eventsection}. 
Based on these selections, the event acceptance efficiency and tracklet acceptance efficiency can be defined. 
Following the Ref. \cite{Belyaev:2020wok}, the event acceptance efficiency $E_{AE}$ is defined as the ratio of the number of the charged Higgs $H^\pm_2$ events passing the event selection criteria ($N_{es}$) to the total number of generated charged Higgs $H^\pm_2$ events ($N$). Similarly, the tracklet acceptance efficiency $T_{AE}$ is defined as the ratio of number of reconstructed events where at least one charged Higgs $H_2^\pm$ is identified ($n_{rec}$) to the total number of charged Higgs $H_2^\pm$ in events that have passed the event selection stage ($n$). For validation of our computation, we used the MSSM framework as used by ATLAS and compared with their reported results as peresented in Appendix~\ref{App:validation}. In Table~\ref{eff_table}, we present the event acceptance efficiency and the tracklet reconstruction efficiency as functions of the lifetime $\tau$. For larger values of $\tau$, the efficiencies are maximized, leading to visible cross sections of order ${\cal O}(10^{-4})$ fb, whereas for smaller $\tau$ the efficiencies decrease rapidly. This loss of efficiency arises because a short-lived (small $\tau$) $H_2^\pm$, typically decays before traversing a sufficient portion of the tracking detector to produce a tracklet that can be reconstructed. 
\begin{table}[h]
\centering
\begin{tabular}{c|c|cccc|cccc}
\hline\hline
\textbf{BPs} & ${\tau~(\text{ns})}$ 
& \multicolumn{4}{c|}{{Event Acceptance Eff.} ($E_{AE}$)} 
& \multicolumn{4}{c}{{Tracklet Acceptance Eff.} ($T_{AE}$)} \\ \cline{3-10}
 &  & $\sqrt{s}=$13 TeV &14 TeV& 27 TeV & 100 TeV & 13 TeV &14 TeV & 27 TeV &100 TeV \\
\hline
 & (a) 0.159  & 0.075&0.0294 & 0.024 & 0.069 &        0.0290 & 0.034 &0.086 & 0.095 \\
\textbf{BP1}   &(b) 0.015 & 0.074& 0.0306& 0.027 & 0.073 &   $-$ &$-$ & $-$ & $0.021$ \\
    &(c) 0.004 & 0.072&0.0296 & 0.015 & 0.051 &      $-$ &$-$& $-$ & 0.008 \\
\hline
 &(a) 0.196  & 0.134 &0.0732& 0.045 & 0.071 &       0.0547 &0.0784 &0.074 & 0.064 \\
\textbf{BP2} &(b) 0.038 & 0.169&0.0873 & 0.033 & 0.059 & $0.0091$ &0.0043 & $0.011$ &  $0.013$ \\
    &(c) 0.013 & 0.089 & 0.0395& 0.022 & 0.066 & $-$ &$-$&  $-$ &  $0.0016$ \\
\hline
 &(a) 0.159  & 0.169& 0.0990 & 0.021 & 0.056 &   0.050 &0.0583& 0.062 & 0.050 \\
\textbf{BP3} &(b) 0.034 & 0.187 & 0.0964&  0.023 & 0.058 &   0.0017 & 0.006& 0.0016 & 0.0018 \\
    &(c) 0.015 & 0.146& 0.0877 & 0.024 & 0.061 &   $-$ &$-$& $-$ & $ 10^{-5}$ \\
\hline \hline
\end{tabular}
\caption{The event acceptance efficiency ($E_{AE}$), and the tracklet 
 acceptance efficiency ($T_{AE}$) at $\sqrt{s}=13,~14,~ 27, ~{\rm and}~100$ TeV for the selected benchmark points in Table~\ref{tb:BPs}. The $"-"$ marks indicate very negligible values, assumed to be zero.}
\label{eff_table}
\end{table}
The total visible cross section for the charged tracks is given in Table \ref{visxs_table}. Notice that for $\sqrt{s}=$13 TeV, these are several orders of magnitude below the current limit of the ATLAS model independent cross section, indicating that such signals are currently beyond the sensitivity of the existing LHC searches. However, these results suggest that future colliders with significantly higher luminosity or improved tracker granularity could probe this region of parameter space, offering potential discovery prospects for long-lived charged scalar scenarios.  
\begin{table}[h]
\centering
\begin{tabular}{c|c|c|c|c}
\hline\hline
&\multicolumn{4}{|c}{Visible cross section $\sigma_{\rm vis}$ (fb)} \\
\cline{2-5}
\textbf{BPs}&$\sqrt{s}=13$ TeV&14 TeV& 27 TeV&100 TeV\\
\hline\hline
 &(a) $7.13\times10^{-5}$ &$4.44\times10^{-5}$& $8.25\times10^{-4}$           & 0.059 \\
\textbf{BP1} &(b)  ~~~~$-$~~~~~~~~ &$-$ & $-$ & $0.011$ \\
    &(c) ~~~~~$-$~~~~~~~~ &$-$  &$-$ & 0.002 \\
\hline
 &(a) $9.01\times10^{-5}$ & $1.07\times10^{-4}$& 0.0013           & 0.051 \\
 \textbf{BP2}&(b) $3.68\times10^{-5}$ & $1.71\times10^{-4}$& $2.65\times10^{-4}$ & 0.014 \\
    &(c)  ~~~~~$-$~~~~~~~~ & $-$ & $-$ & $0.0004$ \\
\hline 
&(a) $4.39\times10^{-5}$  &$5.09\times10^{-5}$ &$3.52\times10^{-4}$         & $8.08\times 10^{-4}$ \\
\textbf{BP3}&(b) $1.23\times10^{-6}$ &$4.74\times10^{-6}$& $8.83\times10^{-6}$ & $8.84\times 10^{-4}$ \\
    &(c)  ~~~~~$-$~~~~~~~~ &$-$&  $-$ &  $4.41\times 10^{-6}$ \\
\hline \hline
\end{tabular}
\caption{Visible cross sections ($\sigma_{\rm vis} = \epsilon~\sigma_{\rm prod}$) at $\sqrt{s}=13$, 14, 27, and 100 TeV hadron colliders for the selected BPs. The total efficiency $\epsilon$ is given by $\epsilon=E_{AE}\times T_{AE} $ }
\label{visxs_table}
\end{table}
We shall now discuss the present experimental constraints on the disappearing track signals by the LHC experiments, and prospects at the future HL-LHC and HE-LHC.
\subsubsection{Compatibility with the present LHC limits}
\begin{figure}[h]
 \centering
    \includegraphics[width=.5\textwidth]{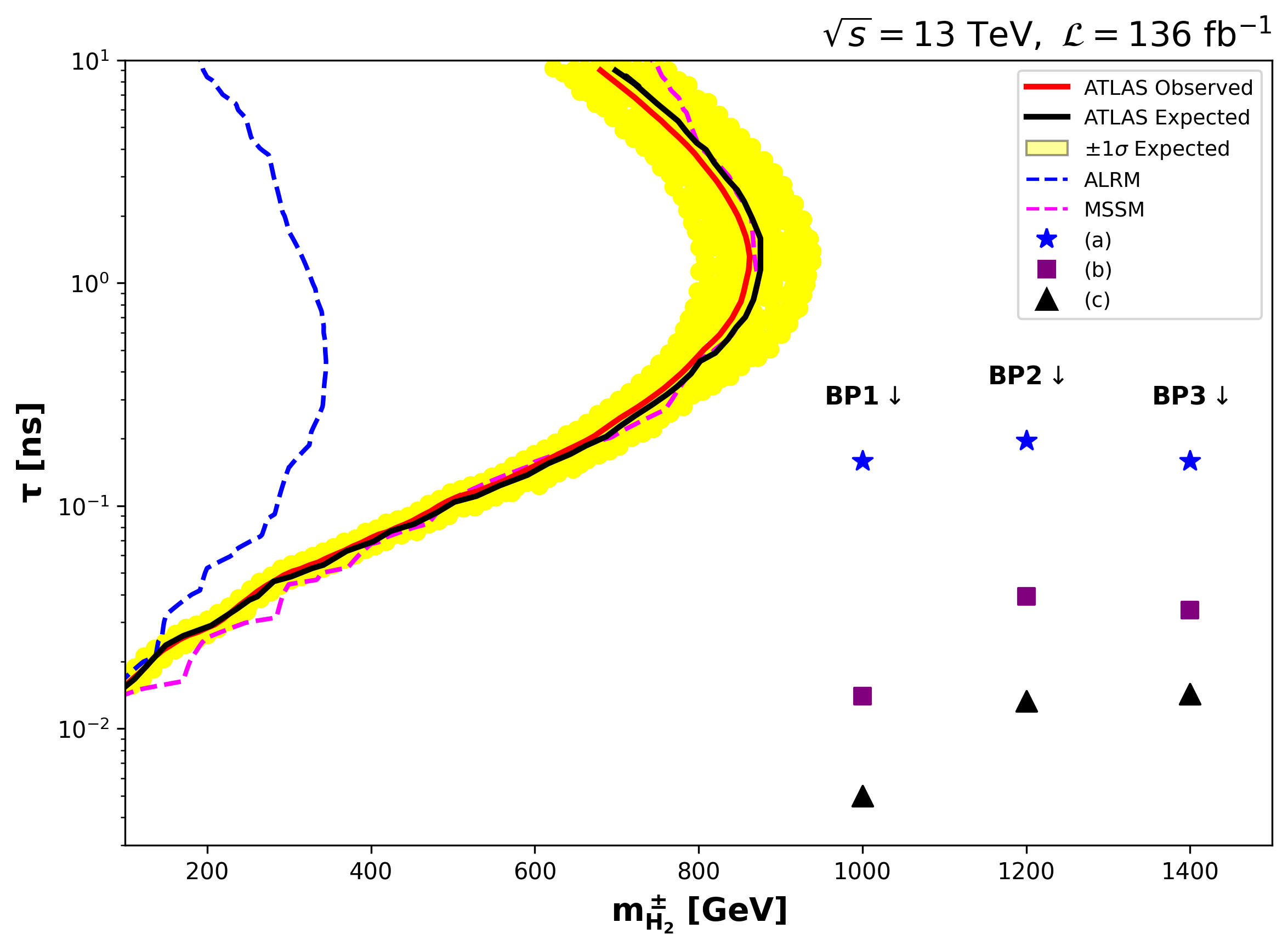}
 \caption{\label{ctau_recast}95\% confidence level limits on the lifetime (in ns) and masss(in GeV) of charged Higgs $H_2^\pm$ adapted from 
ATLAS long-lived chargino searches \cite{ATLAS:2022rme}. 
Dashed-magenta curve is our reconstruction of the corresponding MSSM case studied by ATLAS.
Position of the benchmark points BP1, BP2, and BP3 along with sub-catogories as per Table~\ref{tb:BPs} are indicated. The left side of the curve is excluded at 95\% CL.  The BP's are also consistent with the CMS results \cite{CMS:2020atg, hepdata.95354}. The exclusion boundary in each model is defined by the condition $\sigma_{\rm prod}\times\epsilon \;=\; $
0.037 fb\cite{ATLAS:2022rme}, 
where $\epsilon$ 
denotes the total signal efficiency, as discussed in the text.}
\end{figure}

Both the ATLAS and the CMS have searched for disappearing track signatures and presented their interpretation on the MSSM parameter space. In Fig.~\ref{ctau_recast}, we reproduce the ATLAS interpretation of disappearing track searches at a center-of-mass energy of 
$\sqrt{s}=13$ TeV with an integrated luminosity of ${\cal L}=136$ fb$^{-1}$, as presented for the MSSM in Ref.~\cite{ATLAS:2022rme}. 
The red and black curves denote the observed and expected exclusion limits, respectively, while the yellow band indicates the $\pm 1\sigma$ uncertainty around the expected limit. The blue dashed curve corresponds to the recast exclusion limit obtained for the ALRM considered in this study. The parameter space lying to the left of the exclusion curves is excluded at the 95\% confidence level. The exclusion boundary for a specific model is calculated as $\sigma_{\rm prod}\times \epsilon$, where $\sigma_{\rm prod}$ is the fiducial cross section within the model, after imposing the selection criteria, and $\epsilon=E_{EA}\times T_{AE}$ as defined in Section~\ref{subsec:observability}.
The benchmark points selected for further analysis (Table~\ref{tb:BPs}) are indicated in Fig.~\ref{ctau_recast} as BP1, BP2 and BP3 with the markers stars (a), squares (b) and triangles (c).
The CMS have a similar bounds from their disappearing track searches \cite{CMS:2020atg}. Clearly, all the selected benchmark points lie within the experimentally allowed region of the parameter space. 

\subsubsection{Sensitivity at the future HL-LHC and HE-LHC}
\begin{figure}[H]
 \centering
   \includegraphics[width=.6\textwidth]{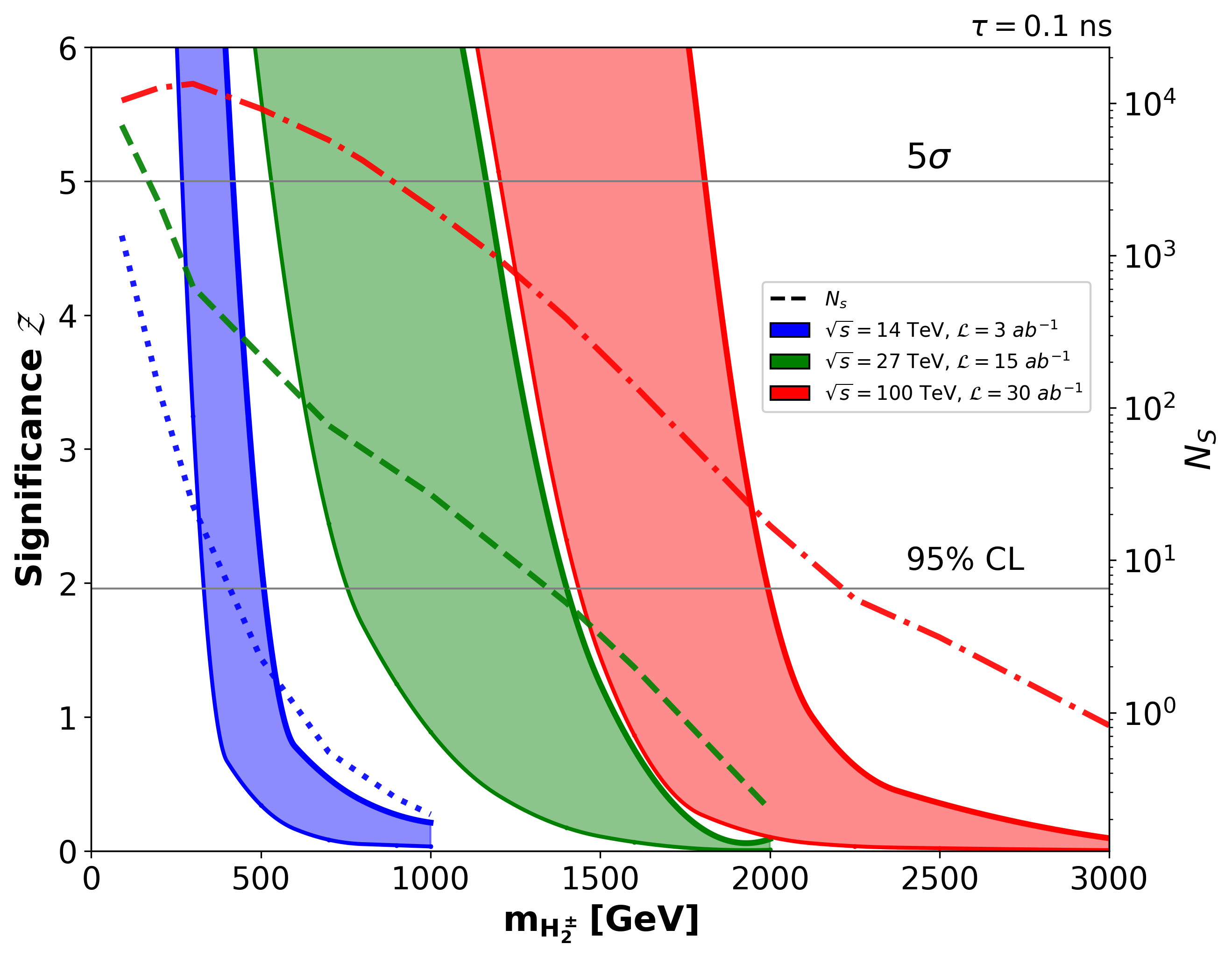}
  \caption{\label{fig:Zband}  Expected significance $\mathcal{Z}$ of the charged Higgs $H_2^\pm$ of life time $\tau=0.1$ ns as a disappearing track overlayed with number of signal events $N_S$ (blue dotted, green dashed and red dash-dotted curves). The HL-LHC at  $\sqrt{s}=14$ TeV and 3000 fb$^{-1}$ luminosity (blue) and the HE-LHC at $27$ (green) and 100 TeV (red) with luminosities of $15$ and $30$ ab$^{-1}$, respectively, are considered.  The horizontal lines  denote the 95\% CL exclusion and $5\sigma$ discovery criteria. 
  }
\end{figure}

\begin{figure}[h]
 \centering
  \includegraphics[width=.46\textwidth]{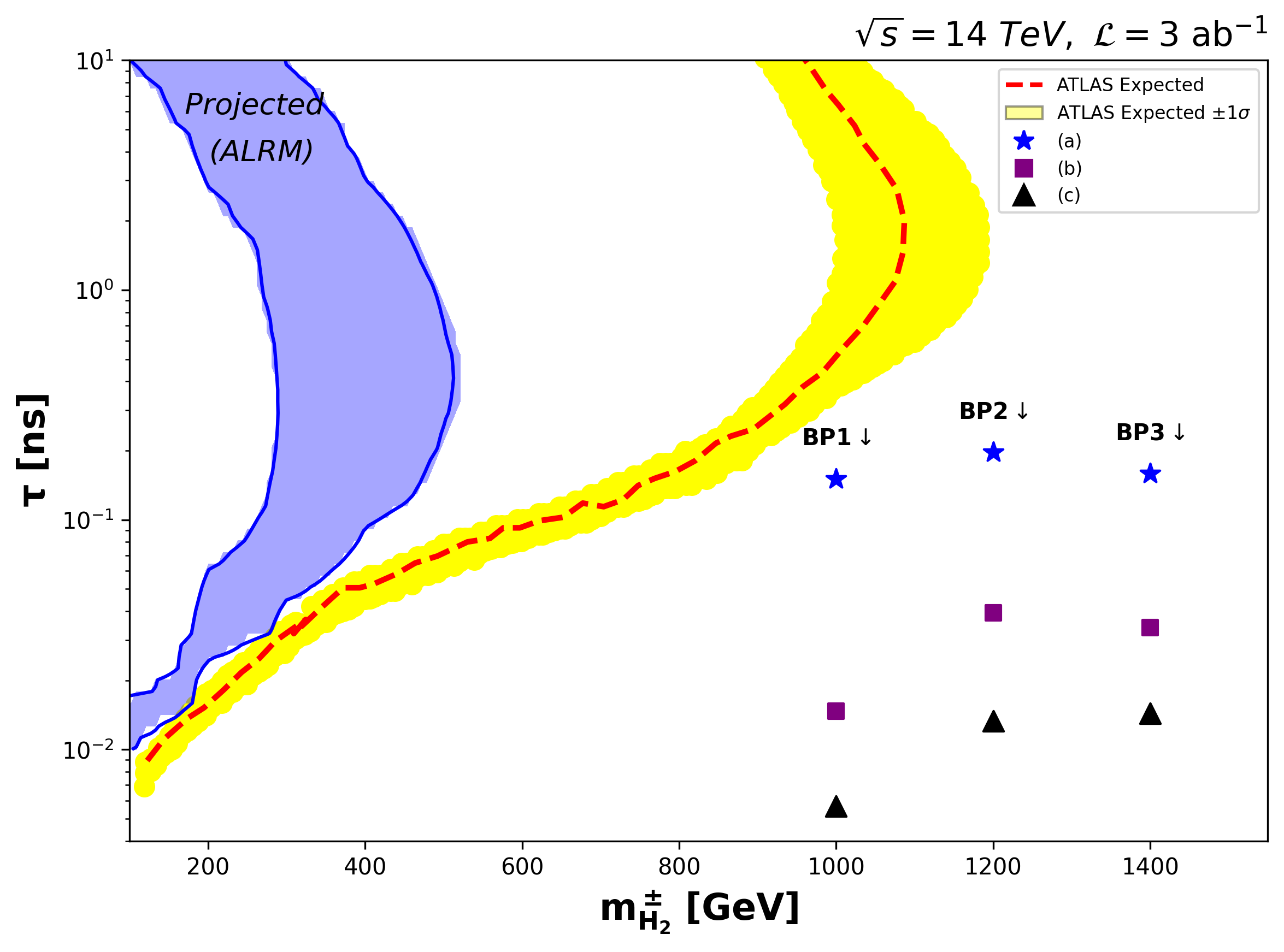}\\
  \includegraphics[width=.46\textwidth]{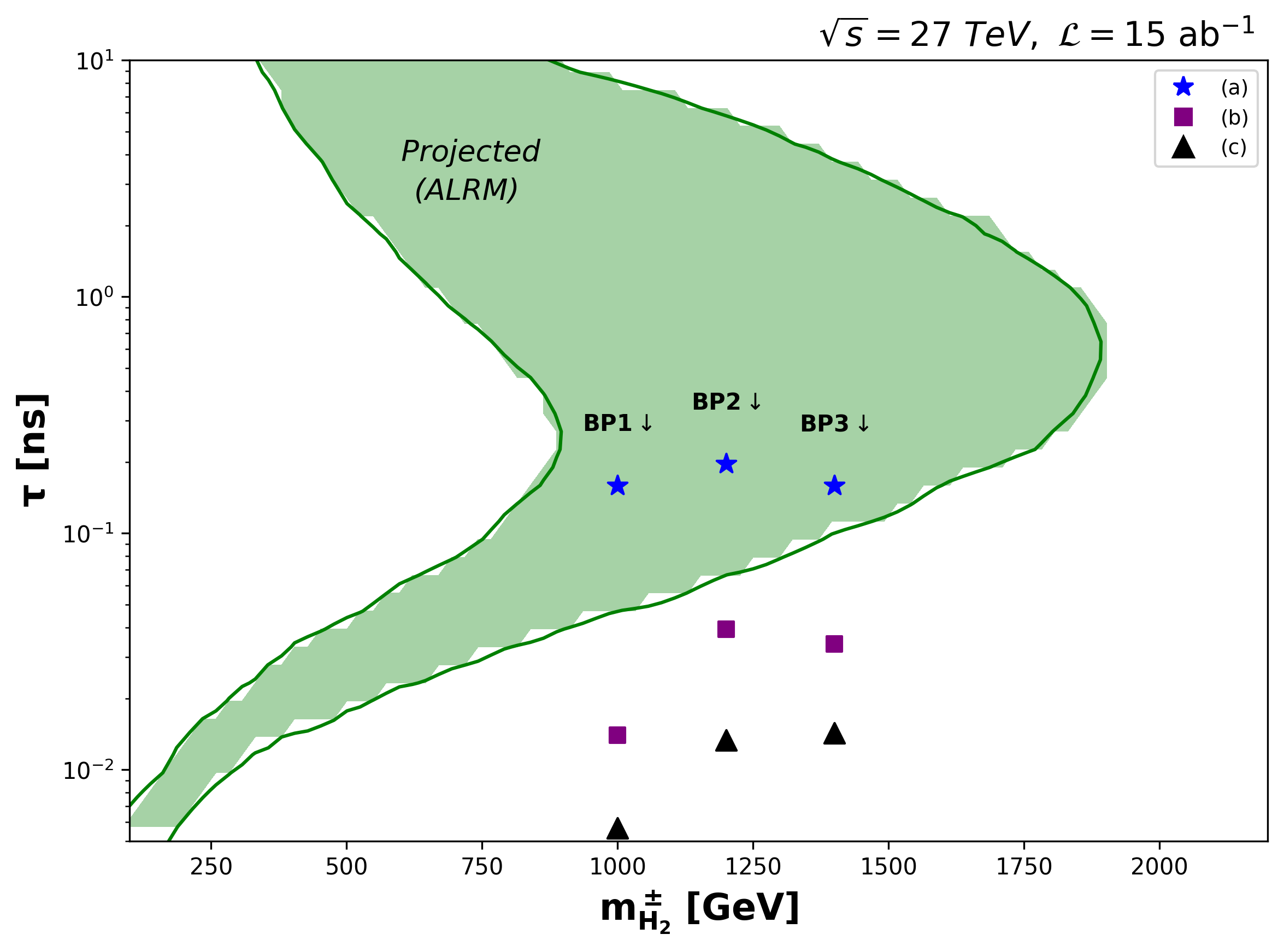}
  \includegraphics[width=.46\textwidth]{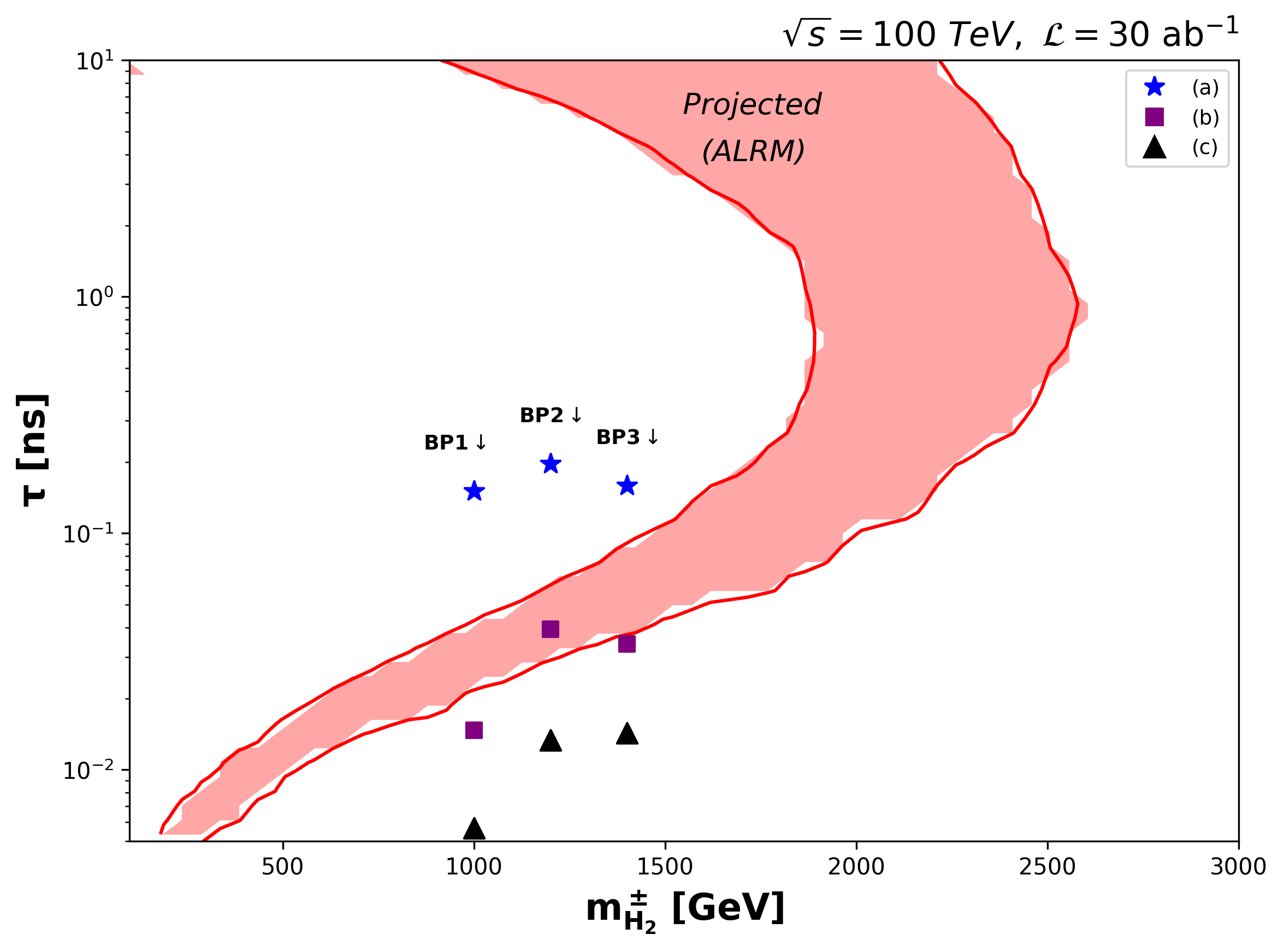}

 \caption{\label{ctau_recastHELHC}95\% confidence level conservative limits on the mass (in GeV) and lifetime (in ns) of charged Higgs $H_2^\pm$ 
at $\sqrt{s}=14$ TeV, $\sqrt{s}=27$ TeV , and $\sqrt{s}=100$ TeV, respectively. The edges of the shaded regions (to the left) represent the exclusion lines with $N_B/5$  and $5 N_B$, respectively. The exclusion contours are based on the values of $\mathcal{Z}$ at 95\% CL.}
\end{figure}

As mentioned earlier,  the SM background rates are very difficult to estimate in the disappearing track analysis. It is because, this backgrounds basically come from the fake tracklets, missing leptons and charged hadrons and are difficult to model accurately by generating the MC events. Therefore, a data driven approach based on an empirical parametrization reported by the ATLAS collaboration\cite{ATLAS:2017oal} is used for a typical background estimate ($N_B$).  
The differential distribution of the fake tracklets (disappearing charged tracks) with their $p_T$  can be parameterized as
\begin{equation}
    \frac{dN_B}{dp_T}=N_0e^{[-p_0\rm log(p_T)-p_1\rm (log(p_T))^2]},
    \label{eq:Byield}
\end{equation}
where the parameters $p_0=0.894$ and $p_1=0.057$ are extracted from a fit to the
fake tracklet data from the 13 TeV LHC. The overall normalization factor $N_0$ is fixed by matching the total number of background events reported by the ATLAS for tracks with $p_T>20$ GeV \cite{ATLAS:2017oal}. For higher energy collider beyond the LHC, the same functional dependence of background on the track momentum is assumed. The overall normalization is rescaled according to the ratio of the production cross sections of $Z+$jets processes at the corresponding collider energy to that at the $13$ TeV LHC. While $Z+$jets processes are dominant, there are also backgrounds from $W+$jets and $t\bar{t}$+jets processes. To account for these additional sources, we allow the background to vary from 20\% to 500\% of its central value, corresponding to a range 
$N_B/5$ to $5N_B$. This procedure provides a conservative assessment of the impact of background uncertainties on the projected reach of the disappearing track searches. At the 27 TeV collider with integrated luminosity of 15 /ab, the projected background yield is around 28 \cite{Han:2018wus}. This is increased moderately to 165 events at the 100 TeV collider with integrated luminosity of 30 /ab. To determine the projected  exclusion limit in the $m_{H_2^\pm}-\tau$ plane, we quantify the significance as
\begin{equation}
    \mathcal{Z}= \frac{N_S}{\sqrt{N_B+(\Delta_BN_B)^2+(\Delta_SN_S)^2}},
\end{equation}
where $N_S$ and $N_B$ correspond to the number of signal events and background events. $\Delta_S$ and $\Delta_B$ are the corresponding  systematic uncertainties. Throughout our analysis, we conservatively assume $\Delta_B=20\%$ and  $\Delta_S=10\%$. In Fig.\ref{fig:Zband}, we plot the significance as a function of $m_{H_2^\pm}$ for a representative value of mean life-time $\tau=0.1$ ns. 
Depending on this $\mathcal{Z}$ values, we derive the exclusion contours in the $m_{H_2^\pm}-\tau$ parameter space, as illustrated in the Fig\ref{ctau_recastHELHC} at $\sqrt{s}=$ 14, 27 and 100 TeV, respectively. The left and the right edges of the shaded regions correspond to  the $\mathcal{Z}$ values with $N_B/5$ and $5 N_B$ (20\% background and 500\% background), respectively. Regions to the left side of these lines correspond to the 95\% CL exclusion limit.

\section{Conclusion}\label{conclusion}

The Alternative Left-Right Model (ALRM), motivated by an underlying unified $E_6$ framework, differs from the conventional left--right symmetric model (LRSM) through a distinct assignment of matter multiplets. In this work, we have investigated the phenomenology of the charged scalar $H_2^\pm$ in the $R$-parity odd sector of the model, emphasizing its direct connection to dark matter dynamics and the possibility of long-lived particle signatures at colliders.

The global $U(1)_S$ symmetry present at high energies is spontaneously broken at the left--right symmetry breaking scale, leaving behind a residual $Z_2$ symmetry. In the presence of an accidental conserved generalized lepton number $L = S + T_{R'3}$, this discrete symmetry effectively plays the role of an $R$-parity--like symmetry, stabilizing the lightest $Z_2$-odd particle and providing a natural dark matter candidate. The charged scalar $H_2^\pm$, being odd under this symmetry, necessarily decays into the dark sector.

We have focused on regions of parameter space consistent with the observed dark matter relic abundance and current direct detection constraints, and in particular considered scenarios featuring a compressed dark sector spectrum, with the mass splitting between $H_2^\pm$ and the scalar dark matter candidate at the sub-GeV scale. Such a compressed spectrum, chosen to explore the phenomenology of long-lived charged scalars, suppresses the decay width of $H_2^\pm$, leaving it long-lived on detector length scales.
The delayed decay of $H_2^\pm$ leads to extremely soft visible final states, giving rise to characteristic disappearing track signatures at the LHC. We identified several benchmark points with charged Higgs masses in the range of $1$--$1.4~\text{TeV}$ whose lifetimes lie within the sensitivity reach of the present and the future collider detectors, at the same time respecting the dark matter observations.

By recasting the ATLAS disappearing track analysis, we evaluated the event acceptance and tracklet reconstruction efficiencies for these benchmark scenarios. Owing to the very small production cross sections, the resulting visible signal rates remain well below the projected sensitivity of the HL-LHC. 
However, presence of such long-lived $H_2^\pm$ could be probed through 
HE-LHC at 27 TeV centre of mass energy or a future $100~\text{TeV}$ proton--proton machine, which could significantly extend the reach  over a broad range of masses and lifetimes.

In summary, the ALRM provides a well-motivated framework in which dark matter stability arises naturally from a residual discrete symmetry originating from the breaking of the global $U(1)_S$ symmetry, with its interpretation as an $R$-parity like symmetry ensured by the conserved generalized lepton number. At the same time, the model offers a natural explanation for parity violation through the spontaneous breaking of the extended gauge symmetry. The emergence of a long-lived charged Higgs boson as a direct consequence of the compressed dark sector spectrum leads to disappearing track signatures at colliders. Our study highlights the importance of searches for long-lived particles as powerful probes of extended Higgs sectors and emphasizes the strong complementarity between dark matter experiments and collider observations in testing unified extensions of the Standard Model.

\section{Acknowledgments}
HD would like to thank the Council of Scientific \& Industrial Research  (CSIR), Govt. of India for the senior research fellowship. PP and Avnish thank the DST, India for financial support with SERB CRG (CRG/2022/002670). PP thanks Department of Theoretical Physics, CERN for hospitality during the final stages of this work.

\appendix
\appendix 

\section{Production Channels for $H_2^\pm$}\label{production_channel}
\hspace{2cm}
\begin{figure}[htbp]
    \centering
    \begingroup 
    \captionsetup{type=figure} 
    
    \makebox[\textwidth][c]{
        \begin{minipage}{0.33\textwidth}
            \centering
            \begin{tikzpicture}[scale=1.1] 
                \begin{feynman}
                    \vertex (i1) at (0, 1.2) {\(\bar{q}\)};
                    \vertex (i2) at (0, -1.2) {\(q\)};
                    \vertex (a) at (0.8, 0); 
                    \vertex (b) at (2.8, 0); 
                    \vertex (f1) at (3.5, 1.2) {\(H_2^+\)};
                    \vertex (f2) at (3.5, -1.2) {\(H_2^-\)};
                    \diagram* {
                        (i1) -- [anti fermion] (a),
                        (i2) -- [fermion] (a),
                        (a) -- [scalar, edge label=\({\text{\small $h/H^0_{2,3}/A_2$}}\), every node/.append style={yshift=0.1cm}] (b),
                        (b) -- [scalar] (f1),
                        (b) -- [scalar] (f2),
                    };
                \end{feynman}
            \end{tikzpicture}
            \caption*{\small (a)}
        \end{minipage}
        \hfill
        \begin{minipage}{0.33\textwidth}
            \centering
            \begin{tikzpicture}[scale=1.1] 
                \begin{feynman}
                    \vertex (i1) at (0, 1.2) {\(\bar{q}\)};
                    \vertex (i2) at (0, -1.2) {\(q\)};
                    \vertex (a) at (0.8, 0); 
                    \vertex (b) at (2.8, 0); 
                    \vertex (f1) at (3.5, 1.2) {\(H_2^+\)};
                    \vertex (f2) at (3.5, -1.2) {\(H_2^-\)};
                    \diagram* {
                        (i1) -- [anti fermion] (a),
                        (i2) -- [fermion] (a),
                        (a) -- [photon, edge label=\({\text{\small $\gamma/Z/Z'$}}\), every node/.append style={yshift=0.1cm}] (b),
                        (b) -- [scalar] (f1),
                        (b) -- [scalar] (f2),
                    };
                \end{feynman}
            \end{tikzpicture}
            \caption*{\small (b)}
        \end{minipage}
        \hfill
        \begin{minipage}{0.33\textwidth}
            \centering
            \begin{tikzpicture}[scale=1.1] 
                \begin{feynman}
                    \vertex (a) at (0, 1.2) {\(\bar{q}\)};
                    \vertex (b) at (0, -1.2) {\(q\)};
                    \vertex (c) at (3.5, 1.2) {\(H_2^-\)}; 
                    \vertex (d) at (3.5, -1.2) {\(H_2^+\)};
                    \vertex (e) at (1.75, 0.7); 
                    \vertex (f) at (1.75, -0.7);
                    \diagram* {
                        (a) -- [anti fermion] (e) -- [scalar] (c),
                        (b) -- [fermion] (f) -- [scalar] (d),
                        (e) -- [anti fermion, edge label=\({\text{\small $d'_i$}}\)] (f),
                    };
                \end{feynman}
            \end{tikzpicture}
            \caption*{\small (c)}
        \end{minipage}        
    }
    \endgroup
    \label{fig:pair_production_large}
\end{figure}


\begin{figure}[htbp]
    \centering
    \begingroup 
    \captionsetup{type=figure} 
    
    \makebox[\textwidth][c]{
        \begin{minipage}{0.48\textwidth}
            \centering
            \begin{tikzpicture}[scale=1.1] 
                \begin{feynman}
                    \vertex (i1) at (0, 1.2) {\(\bar{q}\)};
                    \vertex (i2) at (0, -1.2) {\(q\)};
                    \vertex (a) at (0.8, 0); 
                    \vertex (b) at (3.2, 0); 
                    \vertex (f1) at (4.1, 1.2) {\(H_2^\pm\)};
                    \vertex (f2) at (4.1, -1.2) {\(H_1^0/A_1\)};
                    \diagram* {
                        (i1) -- [anti fermion] (a),
                        (i2) -- [fermion] (a),
                        (a) -- [scalar, edge label=\({\text{\small $H_1^\pm/W_L^\pm$}}\), every node/.append style={yshift=0.1cm}] (b),
                        (b) -- [scalar] (f1),
                        (b) -- [scalar] (f2),
                    };
                \end{feynman}
            \end{tikzpicture}
            \caption*{\small (d)}
        \end{minipage}
        \hfill
        \begin{minipage}{0.48\textwidth}
            \centering
            \begin{tikzpicture}[scale=1.1]
                \begin{feynman}
                    \vertex (a) at (0, 1.2) {\(\bar{q}\)};
                    \vertex (b) at (0, -1.2) {\(q\)};
                    \vertex (c) at (3.5, 1.2) {\(H_1^0/A_1\)}; 
                    \vertex (d) at (3.5, -1.2) {\(H_2^+\)};
                    \vertex (e) at (1.75, 0.7); 
                    \vertex (f) at (1.75, -0.7);
                    \diagram* {
                        (a) -- [anti fermion] (e) -- [scalar] (c),
                        (b) -- [fermion] (f) -- [scalar] (d),
                        (e) -- [anti fermion, edge label=\({\text{\small $d'_i$}}\)] (f),
                    };
                \end{feynman}
            \end{tikzpicture}
            \caption*{\small (e)}
        \end{minipage}        
    }
    \endgroup
\end{figure}

\hspace{2cm}
\begin{figure}[htbp]
    \centering
    \begingroup 
    \captionsetup{type=figure} 
    
    \makebox[\textwidth][c]{
        \begin{minipage}{0.33\textwidth}
            \centering
            \begin{tikzpicture}[scale=1.0] 
                \begin{feynman}
                    \vertex (i1) at (0, 1.2) {\(\bar{q}\)};
                    \vertex (i2) at (0, -1.2) {\(q\)};
                    \vertex (a) at (0.8, 0); 
                    \vertex (b) at (3.2, 0); 
                    \vertex (f1) at (4.2, 1.2) {\(H_2^+\)};
                    \vertex (f2) at (4.2, -1.2) {\(W_R^-\)};
                    \diagram* {
                        (i1) -- [anti fermion] (a),
                        (i2) -- [fermion] (a),
                        (a) -- [scalar, edge label=\({\text{\small $h/H_{2,3}^0/A_2$}}\), every node/.append style={yshift=0.1cm}] (b),
                        (b) -- [scalar] (f1),
                        (b) -- [photon] (f2),
                    };
                \end{feynman}
            \end{tikzpicture}
            \caption*{\small (f)}
        \end{minipage}
        \hfill
        \begin{minipage}{0.33\textwidth}
            \centering
            \begin{tikzpicture}[scale=1.0]
                \begin{feynman}
                    \vertex (i1) at (0, 1.2) {\(\bar{q}\)};
                    \vertex (i2) at (0, -1.2) {\(q\)};
                    \vertex (a) at (0.8, 0); 
                    \vertex (b) at (3.2, 0); 
                    \vertex (f1) at (4.2, 1.2) {\(H_2^+\)};
                    \vertex (f2) at (4.2, -1.2) {\(W_R^-\)};
                    \diagram* {
                        (i1) -- [anti fermion] (a),
                        (i2) -- [fermion] (a),
                        (a) -- [photon, edge label=\({\text{\small $\gamma/Z/Z'$}}\), every node/.append style={yshift=0.1cm}] (b),
                        (b) -- [scalar] (f1),
                        (b) -- [photon] (f2),
                    };
                \end{feynman}
            \end{tikzpicture}
            \caption*{\small (g)}
        \end{minipage}
        \hfill
        \begin{minipage}{0.33\textwidth}
            \centering
            \begin{tikzpicture}[scale=1.0]
                \begin{feynman}
                    \vertex (a) at (0, 1.2) {\(\bar{q}\)};
                    \vertex (b) at (0, -1.2) {\(q\)};
                    \vertex (c) at (3.5, 1.2) {\(W_R^-\)}; 
                    \vertex (d) at (3.5, -1.2) {\(H_2^+\)};
                    \vertex (e) at (1.75, 0.7); 
                    \vertex (f) at (1.75, -0.7);
                    \diagram* {
                        (a) -- [ anti fermion] (e) -- [photon] (c),
                        (b) -- [fermion] (f) -- [scalar] (d),
                        (e) -- [anti fermion, edge label=\({\text{\small $d'_i$}}\)] (f),
                    };
                \end{feynman}
            \end{tikzpicture}
            \caption*{\small (h)}
        \end{minipage}        
    }
    \endgroup

    \caption{Feynman diagrams for the pair production (a, b, c) and the associated production (d, e, f, g, h) of the charged scalar $H_2^+$.}
    \label{fig:associated_WR_large}
\end{figure}

\section{Validation of the LLP-disappearing track codes:}\label{App:validation} 
We have used the ATLAS disappearing track analysis codes as in the Ref.\cite{Belyaev:2020wok} (available on the link: \href{ https://github.com/llprecasting/recastingCodes/tree/main/DisappearingTracks/ATLAS-SUSY-2016-06/}{{\color{blue}https://github.com/llprecasting/recastingCodes/tree/main/DisappearingTracks/ATLAS-SUSY-2016-06/}}). For validation, we used the minimal super-symmetric model (MSSM) from \href{http://hepmdb.soton.ac.uk/hepmdb:1013.0145}{{\color{blue} High Energy Physics Model Data Base (HEPMDB)}}  with the relevant SLHA files from common resources at \href{https://www.hepdata.net/record/ins1641262?version=5}{{\color{blue} the ATLAS data repository}}. Table \ref{app4} shows the 
results of our computation of the event acceptance efficiency ($E_{EA}$) and the tracklet acceptance efficiency ($T_{AE}$) along with the ATLAS reported values as in the Ref. \cite{ATLAS:2017oal} given in the parentheses. 
\begin{table}[H]
\centering
\begin{tabular}{ccccccc}
\hline \hline
chargino mass &Lifetime& \multicolumn{2}{c}{Acceptance Eff.} & \multicolumn{2}{c}{Tracklet Eff.}& $\rm Error$ \\
 $m_{\chi^\pm}$(GeV)& $\tau$ (ns) & &$E_{AE}$  & $ T_{AE}$  &&  $ E_{AE} .T_{AE}$ ($\rm{\%}$) \\
\hline

400 & 0.2 & & 0.116 (0.0927)  &~ 0.0344 (0.0329) & & $11.3\%$\\

600 & 0.2 & & 0.125 (0.126)~  &~ 0.0266 (0.024) & & $9.9\%$ \\

600 & 1.0 & & 0.125 (0.1133)  &~ 0.0939 (0.094) & & $10.2\%$ \\

\hline \hline
\end{tabular}
\caption{Acceptance and efficiency values for events and tracklets for different signal points. The values inside the parentheses are the values reported in the ATLAS search \cite{ATLAS:2017oal}.}
\label{app4}
\end{table}
The percentage error between our computation and that of the ATLAS values are given in the last column.
Our values deviate maximally around 11\%, which is within the acceptable limits.
 The workflow that we follow are detailed in the Section \ref{DTA}.

\section{Event selection criteria:} \label{eventsection}
We present some details of the event selection criteria discussed in Section \ref{subsec:observability}. We refer to  the ATLAS collaboration for the 14 TeV HE-LHC \cite{ATLAS:2018jjf} analysis, while for the HE-LHC analysis we refer to the Ref. \cite{Han:2018wus}.

\subsection{At the HL-LHC: $\sqrt{s}=14$ TeV , $\mathcal{L}=3000~\rm{fb}^{-1}$:}
The event selection criteria are decided based on the distribution of the fake background events and the signal events. 
\begin{figure}[H]
 \centering
 \includegraphics[width=.45\textwidth]{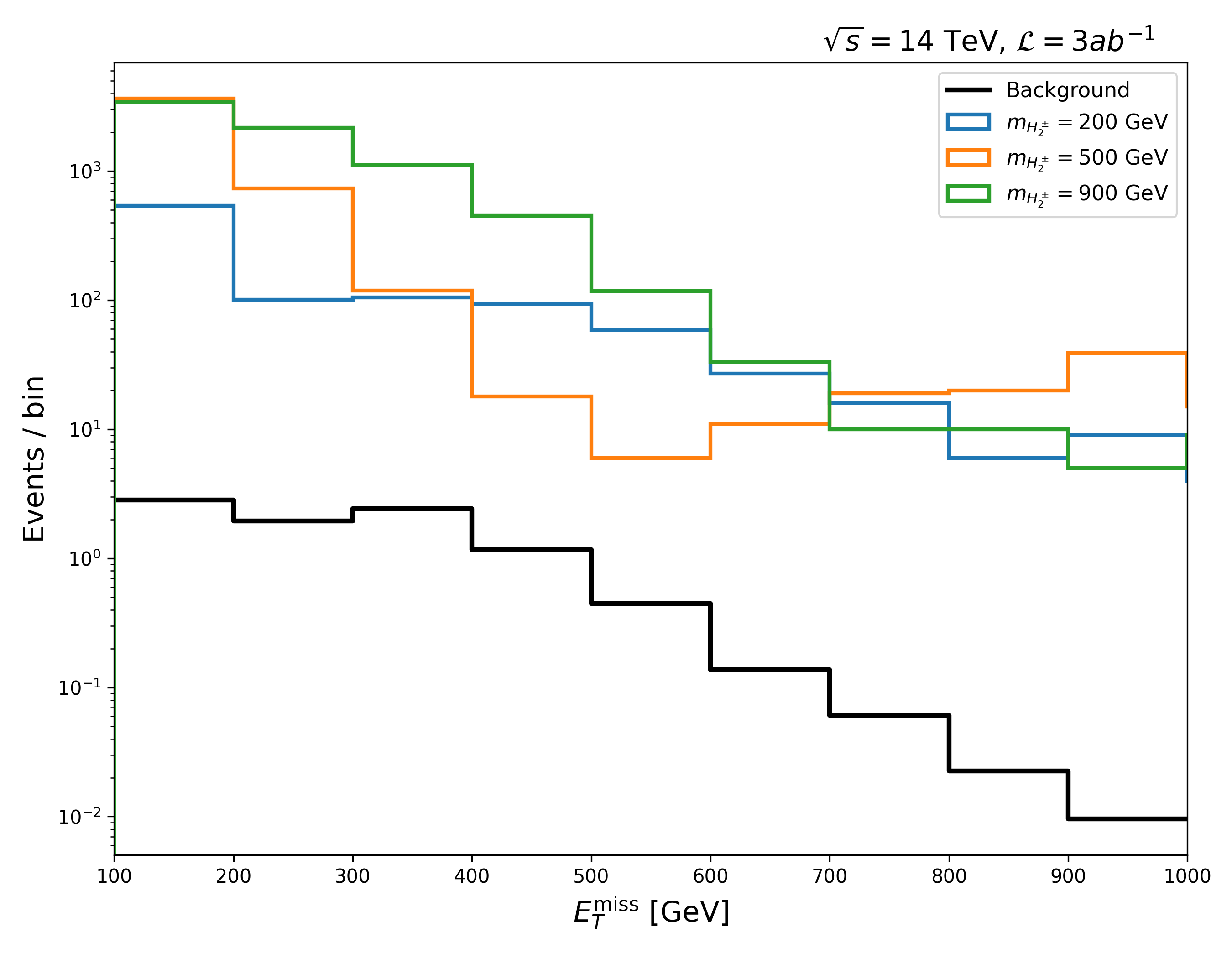} 
 \includegraphics[width=.45\textwidth]{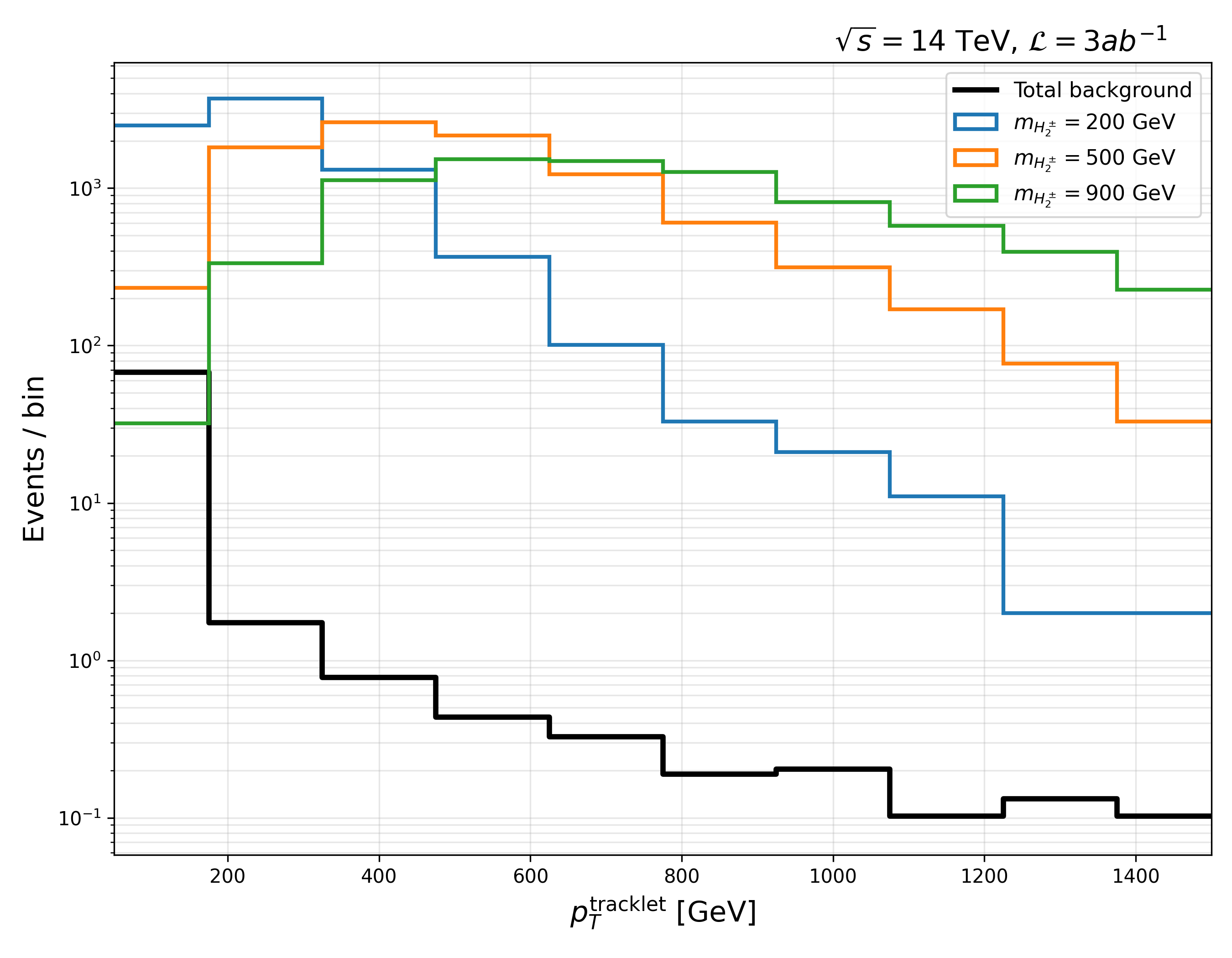}
  \caption{\label{14TeVtest} $\slashed E_T$ distribution (left) and tracklet $p_T$ (right) of $H_2^\pm$ tracklets. The background distributions are from the ATLAS analysis \cite{ATLAS:2018jjf}. }
\end{figure}

 At the HL-LHC, the event selection is performed at the reconstructed level using \texttt{Delphes}. Events are required to satisfy a missing transverse energy trigger of $\slashed{E}_T>110$ GeV. Jets are reconstructed with a requirement of $p_T>20$ GeV and $|\eta|<$ 2.8, while jets used in the azimuthal separation requirement must satisfy $p_T>$ 50 GeV and $|\eta|<$ 2.5. Muons and electrons are selected with $p_T>$ 20 GeV, while $|\eta|<$ 2.8 and $|\eta|<$ 2.47, respectively. To avoid double counting of reconstructed objects, an overlap removal procedure is applied: first, jets are spatially separated from electrons by $\Delta R(\rm{e,jet})>$ 0.2, where $\Delta R =\sqrt{(\Delta\eta)^2+(\Delta \phi)^2}$. This means any jet found within $\Delta R<0.2$ of an electron, the jet is discarded. After this, leptons (both $e$ and $\mu$) are removed if they are found $\Delta R <0.4$ of a jet. Selected events are required to have at least 2 jets, with the leading jet satisfying $p_T(j_1)>$ 300 GeV and final $\slashed{E}_T>$ 300 GeV, and a minimum azimuthal separation min${\Delta\phi(j,\slashed{E}_T)}>1$ for up to four leading jets with $p_T>50$ GeV.

\begin{table}[H]
\centering
\begin{tabular}{c|c}
\hline \hline
Variable & Cuts\\
\hline 
Lepton Veto $p_{T}$ [GeV] & $> 20$\\
min\{$\Delta \phi~(~{\rm jet_{1-4}},\slashed{E}_T )$\} & $> 1$ \\
$\slashed{E}_T $ [GeV] & $> 300$\\
Leading jet $p_T$ [GeV]&$> 300$\\
Leading tracklet $p_T$ [GeV] & $> 150$\\
$\Delta \phi~(~\slashed{E}_T, \rm{trk} )$ &$< 0.5$\\
\hline
\end{tabular}
\caption{Selection criteria for disappearing track signature, as considered by the ATLAS \cite{ATLAS:2018jjf}. }
\label{SR14Tev}
\end{table}
 Tracklet selection is implemented at the generator level and folded with the ATLAS auxiliary efficiency map. Tracklets are required to have $p_T>5$ GeV and $|\eta|<2.2$. Events contributing to the visible cross section must contain a tracklet with max$(p_T)>150$ GeV. The transverse decay radius $R$ is sampled from the Charged Higgs lifetime using the cumulative decay probability, and the tracklet reconstruction efficiency $\epsilon(\eta,R)$ is obtained from the ATLAS efficiency map, which encodes geometrical acceptance, jet isolation, and disappearing condition between the pixel and the SCT layers.
 
\par Since the $H_2^\pm$ decay in the alternative left–right model leads to the same disappearing charged-track signature as in the MSSM wino scenario, we adopt the same set of kinematic observables for event selection. 
\subsection{Section criteria at the HE-LHC:  $\sqrt{s}=27$ TeV  and 
$\sqrt{s}=100$ TeV
}
\par The baseline object definitions and overlap removal criteria are kept identical for all the centre-of-mass energies. This choice provides a conservative recasting framework, while the increased kinematic selections at higher energies as used by the Ref. \cite{Han:2018wus} are presented below.
\begin{table}[H]
\centering
\begin{tabular}{c|c|c}
\hline \hline
Variable & \multicolumn{2}{c}{Selection}\\ \cline{2-3}
 & $\sqrt{s}=27$ TeV&100 TeV\\
\hline 
Lepton Veto $p_{T}$ [GeV] & $> 20$& $> 20$\\
min\{$\Delta \phi~(~{\rm jet_{1-4}},\slashed{E}_T )$\} & $> 1.5 $&$> 1.5$\\
$\slashed{E}_T $ [GeV] & $> 500$&$>1200$\\
Leading jet $p_T$ [GeV]& $ > 400$&$>900$\\
Leading tracklet $p_T$ [GeV] & $> 400$ & $>1000$\\
$\Delta \phi~(~\slashed{E}_T, \rm{trk} )$ &$< 0.5 $&$<0.5$\\
\hline
\end{tabular}
\caption{Selection criteria for disappearing track signature at the HE-LHC following the Ref.\cite{Han:2018wus}. }
\label{SR27Tev}
\end{table}

\section{Tadpole Equation} 
\label{tadpole}
Tadpole equations arising from minimising the scalar potential are 
\begin{align} 
\frac{\partial V}{\partial \phi_{H_1^0}} &= 0\\ 
\frac{\partial V}{\partial \phi_{H_2^0}} &= \frac{1}{\sqrt{2}} \mu_3 v_L v_R  + \lambda_1 v_{u}^{3}  + v_u \Big[(\alpha_{1} + \alpha_{2})(v_{L}^{2} + v_{R}^{2}) - \mu^2_1 \Big]=0\\ 
\frac{\partial V}{\partial \chi_L^0} &=\frac{1}{\sqrt{2}} \mu_3 v_u v_R  + \lambda_{3} v_{L}^{3}  + v_L \Big[(\alpha_{1} + \alpha_{2})v_{u}^{2}  + \lambda_4 v_{R}^{2}  - \mu^2_{2} \Big]=0\\ 
\frac{\partial V}{\partial \chi_R^0} &= \frac{1}{\sqrt{2}} \mu_3 v_u v_L  + v_R \Big[(\alpha_{1} + \alpha_{2} )v_{u}^{2}  + \lambda_{3} v_{R}^{2}  + \lambda_4 v_{L}^{2}  - \mu^2_{2} \Big]=0
\end{align} 
We employ these conditions to rewrite some of the Lagrangian couplings in terms of the remaining couplings and the vevs:
\begin{equation}
    \mu_1^2=(\alpha_{1}+\alpha_2)(v_L^2+v_R^2)+\lambda_1v_u^2+\frac{v_L~v_R~\mu_3}{\sqrt{2}~v_u},
\end{equation}
\begin{equation}
    \mu_2^2=\lambda_3(v_L^2+v_R^2)+(\alpha_{1}+\alpha_2)v_u^2,~~~~~~~~~ ~~~~~
\end{equation}
and
\begin{equation}
\lambda_4=\lambda_3-\frac{\mu_3~v_u}{\sqrt{2}~{v_L~v_R}}.~~~~~~~~~~~~~~~~~~~~~~~~~~~~~
\end{equation}

\printbibliography 

\end{document}